\documentclass[10pt, print, nolongbibliography,groupedaddress,showpacs,showkeys,amssymb,eqsecnum,aps,prd,nofootinbib]{revtex4-2}

\usepackage{microtype}
\usepackage{mathtools, bm}
\usepackage{accents}
\usepackage{tensor}                                                                   

\usepackage{xcolor}                   

\usepackage{hyperref}

\begin{document}                                   

\title{Field redefinition invariant Lagrange multiplier formalism with gauge symmetries}

\begin{abstract}
    It has been shown that by using a Lagrange multiplier field to ensure that the classical equations of motion are satisfied, radiative effects beyond one-loop order are eliminated. It has also been shown that through the contribution of some additional ghost fields, the effective action becomes form invariant under a redefinition of field variables, and furthermore, the usual one-loop results coincide with the quantum corrections obtained from this effective action. In this paper, we consider the consequences of a gauge invariance being present in the classical action. The resulting gauge transformations for the Lagrange multiplier field as well as for the additional ghost fields are found. These gauge transformations result in a set of Faddeev-Popov ghost fields arising in the effective action. If the gauge algebra is closed, we find the Becci-Rouet-Stora-Tyutin (BRST) transformations that leave the effective action invariant. 
\end{abstract}


\author{D. G. C. McKeon}   
\email{dgmckeo2@uwo.ca}
\affiliation{Department of Applied Mathematics, The University of Western Ontario, London, Ontario N6A 5B7, Canada}
\affiliation{Department of Mathematics and Computer Science, Algoma University,
Sault Ste.~Marie, Ontario P6A 2G4, Canada}

\author{F. T. Brandt}  
\email{fbrandt@usp.br}
\affiliation{Instituto de F\'{\i}sica, Universidade de S\~ao Paulo, S\~ao Paulo, SP 05508-090, Brazil}

\author{S. Martins-Filho}   
\email{sergiomartinsfilho@usp.br}
\affiliation{Instituto de F\'{\i}sica, Universidade de S\~ao Paulo, S\~ao Paulo, SP 05508-090, Brazil}

\date{\today}

\maketitle

\section{Introduction}\label{section:intro}

When one quantizes gauge theories such as electrodynamics, Yang-Mills (YM) theory, and the Einstein-Hilbert (EH) theory of gravity, special care must be taken to cancel quantum effects that arise from unphysical gauge fields. Special ``ghost'' fields have been found that perform this function \cite{Faddeev:1967fc, DeWitt:1967ub, Mandelstam:1968hz}. Even after these ghost fields have been introduced, the resulting effective action gives rise to divergent quantum effects that are removed through renormalizing the quantities that characterize the classical theory. This program works well in electrodynamics \cite{Dyson:1949bp} and YM theory \cite{tHooft:1971akt, tHooft:1971qjg}. With the EH action these divergences vanish on mass-shell at one loop order, but do not disappear at two loop order when the equation of motion are satisfied \cite{Goroff:1985th, vandeVen:1991gw}.

Many different approaches, such as supergravity \cite{freedman:2012}, string theory \cite{west:2012}, loop gravity \cite{Rovelli:1997yv}, and asymptotic safety \cite{Bednyakov:2023fmc}, have been proposed to cure this problem with quantum gravity. Perhaps the simplest approach is to introduce a Lagrange multiplier  field that ensures that classical equation of motion for the gravitational field is satisfied \cite{Brandt:2018lbe, Brandt:2019ymg} as then all radiative effects beyond one-loop order are absent \cite{McKeon:1992rq} with the remaining one-loop effects being twice these one-loop effects coming from the usual quantization procedure. This approach can be applied consistently to gauge theories coupled to matter \cite{Brandt:2021qgh} or in first order form \cite{Brandt:2020gms}. In Ref. \cite{Brandt:2021nev}, the thermal effects on gauge theory supplemented with Lagrange multiplier fields were investigated.
Another example of a field theory in which a Lagrange Multiplier field is used to impose a constraint and is involved in radiative corrections is provided by Ref. \cite{Alvarez:2023kuw}.

A significant improvement on this approach involves the introduction of a functional determinant into the measure of the path integral that has the effect of rendering the path integral invariant under a field redefinition.  It also results in the quantum effects in this approach being exactly equal to the one-loop effects normally encountered. 
(The factor of two mentioned above that occurs when only using a Lagrange multiplier field no longer arises when this functional determinant also is introduced.) \cite{Brandt:2022kjo}

This functional determinant can be incorporated into the effective action through the introduction of a pair of Fermionic ghosts and a single Bosonic ghost field. These ghosts are similar to Lee-Yang ghost fields \cite{Bastianelli:1998jb}, which appear in the context of the worldline formalism \cite{Schubert:2001he, Edwards:2021elz}. In this paper, we consider the consequence of there being a gauge invariance in the classical Lagrangian. It is shown that the Lagrange multiplier field and the ghost fields all participate in this gauge transformation, with additional gauge invariances occurring in the effective Lagrangian. 

We then show that if these gauge transformations are closed, then a BRST transformation \cite{Becchi:1974xu, Tyutin:1975qk} can be defined which leaves the effective Lagrangian invariant. Such an invariance leads to the full theory being unitary \cite{Kugo:1977zq} and renormalizable \cite{Baulieu:1983tg}.

This paper is organized as follows. In Section II, we review the field redefinition invariant Lagrange multiplier formalism for non-singular classical actions. The case of a classical action with a gauge symmetry is considered in Section III and we show that additional gauge invariances appear. We then quantize by using an extension of the Faddeev-Popov procedure \cite{Faddeev:1967fc}. Then, in Section IV, we derive the corresponding BRST transformations in the framework of the extended Lagrange multiplier formalism. In Section V, we consider YM theory, a non-linear gauge theory, to demonstrate this formalism. We also find the corresponding BRST transformation. In appendix A, we illustrate the field redefinition invariance of path integrals when using the extended Lagrange multiplier formalism. 
In appendix B, we briefly review some of the material of references \cite{Brandt:2018lbe, Brandt:2019ymg, McKeon:1992rq,Brandt:2021qgh, Brandt:2020gms, Brandt:2021nev}, to show how the Lagrange multiplier field eliminates loop effects beyond one loop order and that no one-particle irreducible contributions involving two or more Lagrange multiplier fields occur at one-loop. The propagators of the Yang-Mills theory are also derived. 
In appendix C, we study the gauge algebra of YM type gauge theories in the extended Lagrange multiplier theory. A superdeterminant identity is derived in detail in the appendix D. In appendix E, we obtain the Zinn-Justin master equation \cite{Zinn-Justin:1974ggz} associated with the BRST transformation derived in Section IV. The relation with the Batalin-Vilkovisky formalism \cite{Batalin:1981jr, Batalin:1983ggl} is briefly discussed.

\section{The extended Lagrange multiplier theory}\label{section:EXLagrange multiplierT}

If one has a set of classical fields $ \phi_{i} (x) $ whose classical Lagrangian is $ \mathcal{L}_{\text{cl}} ( \phi_{i} ) $, then a Lagrangian multiplier field $ \lambda_{i} (x) $ can be used to ensure that the classical equation of motion 
\begin{equation}\label{eq:1}
    \mathcal{L}_{\text{cl} , i} \equiv \frac{\partial \mathcal{L}_{\text{cl}}}{\partial \phi_{i}} =0
\end{equation}
is obeyed. The path integral 
\begin{equation}\label{eq:2}
    I = \int \mathop{\mathcal{D} \phi_i} \mathop{\mathcal{D} \lambda_{i}} \exp{ i \int \mathop{d^{}x}} \left ( \mathcal{L}_{\text{cl}} ( \phi_{i} ) + \lambda_{i} \mathcal{L}_{\text{cl} , i} ( \phi_{i} )\right )
\end{equation}
can in principle be evaluated using the functional analogue of the usual results 
\begin{subequations} \label{eq:3}
    \begin{align}
    \label{eq:3a}
    \int_{- \infty}^{\infty} \frac{ \mathop{d \lambda}}{2 \pi} \exp \left( i \lambda f (x)\right) ={}&\mathop{\delta} \left ( f (x)\right )  
  \\  \intertext{and}
    \label{eq:3b}
    \int_{- \infty}^{\infty} \exp( i I[x]) \mathop{\delta} \left ( f(x)\right ) ={}& \sum_{i}^{} \exp \left ( i I[x_{i}] \right ) | f'(x_{i} ) |^{-1},
\end{align}
\end{subequations}
where $ f( x_{i} ) = 0$. It follows that \cite{Brandt:2021qgh} 
\begin{equation}\label{eq:4}
    I = \sum_{i}^{} \exp \left ( i \int \mathop{d^{}x} \mathcal{L}_{\text{cl}} ( \bar{\phi}_{i} (x))\right ) \left [ \det \mathcal{L}_{\text{cl} , ij} ( \bar{\phi}_{i} )\right ]^{-1},
\end{equation}
where 
\begin{equation}\label{eq:5}
    \mathcal{L}_{\text{cl} , i}  \left ( \bar{\phi}_{ i} (x)\right ) =0.
\end{equation}
In Eq.~\eqref{eq:4}, the exponential is the sum of all tree-level diagrams arising from $ \mathcal{L}_{\text{cl}} ( \phi_{i} ) $ alone \cite{Boulware:1968zz}. The determinant is the square of the determinant coming from all one-loop diagrams and hence all one-loop results that normally arise acquire an extra factor of two. No contributions corresponding to higher-loop diagrams occur in the exact result of Eq.~\eqref{eq:4}.

In Ref. \cite{Brandt:2022kjo}, a further modification of the functional integral in Eq.~\eqref{eq:2} is introduced.  The functional measure in Eq.~\eqref{eq:2} is supplemented by the functional determinant 
\begin{equation}\label{eq:6}
    \det\nolimits^{1/2} [ \mathcal{L}_{\text{cl} , ij} ( \phi_{i} )].
\end{equation}
This is shown to leave $I$ invariant under a field redefinition.\footnote{See the appendix A for an illustration of how this comes about.} It is also apparent from combing Eqs.~\eqref{eq:4} and \eqref{eq:6} that all quantum effects now reduce to the one-loop effects arising from $ \mathcal{L}_{\text{cl}} ( \phi_{i} )$ alone.

We can exponentiate the functional determinant of Eq.~\eqref{eq:6} through use of a Bosonic ghost field $ \chi_{i} (x) $ and a pair of Fermionic ghost fields $ \psi_{i} (x)$ and $ \theta_{i} (x)$ \cite{berezin:1987} 
\begin{equation}\label{eq:7}
    \det\nolimits^{1/2} \mathcal{L}_{\text{cl} , ij} ( \phi_{i} ) = \int \mathop{\mathcal{D} \chi_{i}} \mathop{\mathcal{D} \psi_{i}} \mathop{\mathcal{D} \theta_{i}}  \exp{i \int \mathop{d^{}x} \left [ \mathcal{L}_{\text{cl}, ij} ( \phi_{i} (x)) \left ( \psi_{i} (x) \theta_{j} (x) + \frac{1}{2} \chi_{i} (x) \chi_{j} (x) \right )\right ]}.
\end{equation}
The full classical effective Lagrangian is now 
\begin{equation}\label{eq:8}
    \mathcal{L}_{\text{eff}} ( \phi_{i} , \lambda_{i} , \chi_{i} , \psi_{i} , \theta_{i} ) = 
    \mathcal{L}_{\text{cl}} ( \phi_{i} ) + \lambda_{i} \mathcal{L}_{\text{cl} , i} + \left ( \psi_{i} \theta_{j} + \frac{1}{2} \chi_{i} \chi_{j}\right ) \mathcal{L}_{\text{cl} ,ij} .
\end{equation}
If the classical action\footnote{Let us assume that $ \phi_{i} $ is a real Bosonic field. Note that, the fields $ \lambda_{i} $ and $ \chi_{i} $ must be of the same kind.}, 
\begin{equation}\label{eq:9II}
    S_{\text{cl}} = \int \mathop{d x} \mathcal{L}_{\text{cl}} ( \phi_{i} )
\end{equation}
is non-singular, then the quantization of the Lagrangian \eqref{eq:8} can be realized through the standard path integral procedure \cite{Brandt:2022kjo} 
\begin{equation}\label{eq:10II}
    Z[ \bm{J} ] =
    \int \mathop{\mathcal{D} \bm{\phi_{i}}} \exp i \int \mathop{d x} \left(
    \mathcal{L}_{\text{cl}} ( \phi_{i} ) + \lambda_{i} \mathcal{L}_{\text{cl} , i} + \left ( \psi_{i} \theta_{j} + \frac{1}{2} \chi_{i} \chi_{j}\right ) \mathcal{L}_{\text{cl} ,ij} 
+ \bar{\bm{J}}_{i} \bm{\phi}_{i}  \right), 
\end{equation}
where we used a compact notation in which $ \bm{\phi}_{i} = ( \phi_{i} , \lambda_{i} , \chi_{i} , \psi_{i} , \theta_{i} )$, $ \mathop{\mathcal{D} \bm{\phi}_{i} } \equiv  \mathop{\mathcal{D} \phi_{i}} \mathop{\mathcal{D} \lambda_{i}} \mathop{\mathcal{D} \chi_{i}} \mathop{\mathcal{D} \psi_{i}} \mathop{\mathcal{D} \theta_{i} }$. We also included a source $ \bar{\bm{J}}_{i} \equiv  ( J_{i} , J_{i} + K_{i}   , L_{i}, \bar{\eta}_{i} , \bar{\kappa}_{i} )$, where $ J_{i} $, $ K_{i} $, $ L_{i} $ are ordinary sources and $ \bar{\eta}_{i} $, $ \bar{\kappa}_{i} $ are fermionic sources. Thus, the expanded form of the source term in Eq.~\eqref{eq:10II} is given by
\begin{equation}\label{eq:11II}
    \bar{\bm{J}}_{i} \bm{\phi}_{i} = J_{i} (\phi_{i}  + \lambda_{i} ) + K_{i} \lambda_{i} + L_{i} \chi_{i} + \bar{\eta}_{i} \psi_{i} + \bar{\kappa}_{i} \theta_{i} .
\end{equation}

\section{Gauge Invariance}\label{section:GI}

We now consider the consequences of $ \mathcal{L}_{\text{cl}} ( \phi_{i} ) $ being invariant under the local infinitesimal transformation 
\begin{equation}\label{eq:9}
    \phi_{i} \to \phi_{i} ' = \phi_{i} + H_{ij} ( \phi_{i} ) \xi_{j}
\end{equation}
so that 
\begin{equation}\label{eq:10}
    \mathcal{L}_{\text{cl}} ( \phi_{i} ) = \mathcal{L}_{\text{cl}} ( \phi ') = \mathcal{L}_{\text{cl}} ( \phi_{i} ) + \mathcal{L}_{\text{cl} ,i} H_{ij} \xi_{j} .
\end{equation}

From Eq.~\eqref{eq:10} it follows that 
\begin{equation}\label{eq:11}
    \mathcal{L}_{\text{cl} , i} H_{ij} \xi_{j} =0
\end{equation}
and so immediately it follows that $ \mathcal{L}_{\text{eff}} $ in Eq.~\eqref{eq:8} is also invariant under the local infinitesimal gauge transformation 
\begin{equation}\label{eq:12}
    \lambda_{i} \to \lambda_{i} ' + H_{ij} ( \phi_{i} ) \zeta_{j} .
\end{equation}

From Eq.~\eqref{eq:9} it follows that 
\begin{equation}\label{eq:13}
    \frac{\partial \mathcal{L}_{\text{cl}}}{\partial \phi_{i}} = \frac{\partial \phi_{j} '}{\partial \phi_{i}} \frac{\partial \mathcal{L}_{\text{cl}}}{\partial \phi_{j} '} = \left ( \delta_{ji} + \frac{\partial H_{jk}}{\partial \phi_{i}} \xi_{k}\right ) \frac{\partial \mathcal{L}_{\text{cl}}}{\partial \phi_{j} '} 
\end{equation}
and consequently that 
\begin{equation}\label{eq:14}
    \frac{\partial^{2} \mathcal{L}_{\text{cl}}}{\partial \phi_{i} \partial \phi_{j}} = \frac{\partial \phi_{m} '}{\partial \phi_{i}} \frac{\partial}{\partial \phi_{m} '} \left [ \frac{\partial \mathcal{L}_{\text{cl}}}{\partial \phi_{j} '} + \frac{\partial H_{nk}}{\partial \phi_{j}} \xi_{k} \frac{\partial \mathcal{L}_{\text{cl}}}{\partial \phi_{n} '}\right ]
\end{equation}
which to leading order in $ \xi_{i} $ reduces to 
\begin{equation}\label{eq:15}
    \frac{\partial^{2} \mathcal{L}_{\text{cl}}}{\partial \phi_{i} \partial \phi_{j}} =
    \frac{\partial^{2} \mathcal{L}_{\text{cl}}}{\partial \phi_{i} ' \partial \phi_{j} '} + \left ( \frac{\partial^{2} H_{lk}}{\partial \phi_{i} \partial \phi_{j}} \xi_{k}\right ) \frac{\partial \mathcal{L}_{\text{cl}}}{\partial \phi_{l} '} + 
    \left ( \frac{\partial H_{lk}}{\partial \phi_{i}} \frac{\partial^{2} \mathcal{L}_{cl}}{\partial \phi_{j} ' \partial \phi_{l} '}  + \frac{\partial H_{lk}}{\partial \phi_{j}} \frac{\partial^{2} \mathcal{L}_{cl}}{\partial \phi_{i} ' \partial \phi_{l} '}\right ) \xi_{k} .
\end{equation}
Insertion of Eqs.~\eqref{eq:9}, \eqref{eq:13} and \eqref{eq:15} into Eq.~\eqref{eq:8} and collecting terms dependent on $ \partial \mathcal{L}_{\text{cl}} / \partial \phi_{i} '$ and $ \partial^{2} \mathcal{L}_{\text{cl}} / \partial \phi_{i} ' \partial \phi_{j} ' $ leads to invariance of $ \mathcal{L}_{\text{eff}} $ under the transformations  
 \begin{subequations}\label{eq:16}
 \begin{align}
 \lambda_{i} \to \lambda_{i} ' ={}& \lambda_{i} + \left [ \frac{\partial H_{ik}}{\partial \phi_{j}} \lambda_{j} + \frac{\partial^{2} H_{ik}}{\partial \phi_{m} \partial \phi_{n}}\left ( \psi_{m} \theta_{n} + \frac{1}{2} \chi_{m} \chi_{n}\right )\right ] \xi_{k} , \\
 \psi_{i} \to \psi_{i} ' ={}& \psi_{i} + \frac{\partial H_{ik}}{\partial \phi_{j}} \psi_{j} \xi_{k}, \\
 \theta_{i} \to \theta_{i} ' ={}& \theta_{i}  
+ \frac{\partial H_{ik}}{\partial \phi_{j}} \theta_{j} \xi_{k},\\
 \chi_{i} \to \chi_{i} ' ={}& \chi_{i} 
+ \frac{\partial H_{ik}}{\partial \phi_{j}} \chi_{j} \xi_{k}.
     \end{align}
 \end{subequations} 
In addition, there are also the gauge transformations
 \begin{subequations}\label{eq:17}
    \begin{align}
 \label{eq:17a}
 \lambda_{i} \to \lambda_{i} ' ={}&\lambda_{i}  + H_{ik, j} \left ( \chi_{j} \sigma_{k} + \psi_{j} \tau_{k} - \theta_{j} \pi_{k} \right ), \\ 
     \label{eq:17b}
        \psi_{i} \to \psi_{i} ' ={}& \psi_{i} + H_{ij} \pi_{j}, \\
 \label{eq:17c}
        \theta_{i} \to \theta_{i} ' ={}&\theta_{i} + H_{ij} \tau_{j} ,\\
 \label{eq:17d}
        \chi_{i} \to \chi_{i} ' ={}& \chi_{i} + H_{ij} \sigma_{j} ;
     \end{align}
 \end{subequations}
 where $ \pi_{i} $ and $ \tau_{i} $ are Fermionic gauge functions and $ \sigma_{i} $ is a Bosonic gauge function. These follow from: 
 \begin{equation}\label{eq:18}
     ( \mathcal{L}_{\text{cl} , i} H_{ik} )_{,j} =0
 \end{equation}
 which is a consequence of Eq.~\eqref{eq:11}.

 We now will introduce a gauge fixing Lagrangian 
 \begin{equation}\label{eq:19}
     \mathcal{L}_{\text{gf}} = - \frac{1}{2 \alpha} ( F_{ij} \phi_{j} )^{2} - \frac{1}{\alpha} ( F_{im} \lambda_{m} ) ( F_{i n} \phi_{n} ) - \frac{1}{2 \alpha } ( F_{ij} \chi_{j} )^{2} - \frac{1}{\alpha} ( F_{im} \psi_{m} )( F_{in} \theta_{n} ) .
 \end{equation}
With Bosonic Nakanishi-Lautrup fields \cite{Nakanishi:1966zz, lautrup:1967} $ B_{i} $, $ E_{i} $ and $ G_{i} $ and similarly the Fermionic fields $ \Omega_{i} $ and $ \Xi_{i} $, this can be written as 
 \begin{equation}\label{eq:20}
     \mathcal{L}_{\text{gf}} = \frac{\alpha}{2} \left ( - E_{i}^{2} + 2 B_{i} E_{i} +G_{i}^{2} - 2 \Xi_{i} \Omega_{i}\right ) - E_{i} ( F_{ij} \lambda_{j} ) - B_{i} ( F_{ij} \phi_{j} ) - G_{i} ( F_{ij} \chi_{j} ) - \Xi_{i} (F_{ij} \theta_{j} ) - \Omega_{i} ( F_{ij} \psi_{j} ).
 \end{equation}
 
 Having fixed the gauge invariance of Eqs.~\eqref{eq:9}, \eqref{eq:12}, \eqref{eq:16} and \eqref{eq:17} in this way, we now introduce the Faddeev-Popov ghost contribution to the path integral 
 \begin{equation}\label{eq:21}
     I = \int \mathop{\mathcal{D} \phi_{i}} \mathop{\mathcal{D} \lambda_{i}} \mathop{\mathcal{D} \chi_{i}} \mathop{\mathcal{D} \psi_{i} \mathop{\mathcal{D} \theta_{i}}} \exp{i \int \mathop{d^{}x} \left [ \mathcal{L}_{\text{cl}} ( \phi_{i} ) + \mathcal{L}_{\text{cl} , i} \lambda_{i} + \mathcal{L}_{\text{cl} , ij} ( \phi_{i} ) \left( \psi_{ i} \theta_{j} + \frac{1}{2} \chi_{i} \chi_{j}\right)\right ]} .
 \end{equation}
 This involves first inserting the constant 
 \begin{equation}\label{eq:22}
     \int \mathop{\mathcal{D} \xi_{i}} \mathop{\mathcal{D} \zeta_{i}} \mathop{\mathcal{D} \sigma_{i}} \mathop{\mathcal{D} \pi_{i}} \mathop{\mathcal{D} \tau_{i}}   
     \delta \left \{ F_{ij} \left [ \begin{pmatrix}
                 \phi_{j} \\ 
                 \lambda_{j} \\
                 \chi_{j} \\
                 \psi_{j} \\
                 \theta_{j} 
             \end{pmatrix}
             + 
\begin{pmatrix}
     0 & H_{jp} & 0 & 0 & 0 \\ 
     H_{jp} & X_{jp}  & H_{jp,k} \chi_{k}  & -H_{jp,k} \theta_{k} &H_{jp,k} {\psi}_{k}  \\
     0 & H_{jp,k} \chi_{k} & H_{jp} & 0 & 0 \\
     0 & H_{jp,k} \psi_{k} & 0 & H_{jp} & 0 \\
     0 & H_{jp,k} \theta_{k} & 0 & 0 & H_{jp} \\ 
     \end{pmatrix}
\begin{pmatrix}
    \zeta_{p} \\ 
    \xi_{p} \\
    \sigma_{p} \\
    \pi_{p} \\
    \tau_{p} 
\end{pmatrix}
\right ] - \begin{pmatrix}
    p_{i} \\
    q_{i} \\
    r_{i} \\
    s_{i} \\
    t_{i}
\end{pmatrix}
\right \} \mathop{\rm{Sdet} } \undertilde{M}_{jp}, 
 \end{equation}
where \begin{equation}\label{eq:23} X_{jp} = H_{jp, k} \lambda_{k} + H_{jp,mn} \left( \psi_{m} \theta_{n} + \frac{1}{2} \chi_{m} \chi_{n} \right), \end{equation} $ \undertilde{M}_{jp}$ is the $5 \times 5$ matrix appearing in the argument of  the $ \delta $-function in Eq.~\eqref{eq:22} and $ \mathop{\rm Sdet} \undertilde{M}_{jp}$ is the Faddeev-Popov superdeterminant.

 We then perform the gauge transformations of Eqs.~\eqref{eq:9}, \eqref{eq:12}, \eqref{eq:16}, \eqref{eq:17} with $ (- \xi_{i} , - \zeta_{i} , - \sigma_{i} , - \pi_{i} , - \tau_{i} )$ and then insert the constant 
 \begin{equation}\label{eq:24}
     \int \mathop{\mathcal{D} p_{i}} \mathop{\mathcal{D} q_{i}} \mathop{\mathcal{D} r_{i}} \mathop{\mathcal{D} s_{i}} \mathop{\mathcal{D} t_{i}} \exp i \int \mathop{d^{}x} \left ( - \frac{1}{2 \alpha} p_{i}^{2} - \frac{1}{\alpha} p_{i} q_{i} - \frac{1}{2 \alpha} r_{i}^{2} - \frac{1}{\alpha} s_{i} t_{i}\right ). 
 \end{equation}
 The integral over $ ( \xi_{i} , \zeta_{i}, \sigma_{i} , \pi_{i} , \tau_{i} ) $ are now innocuous multiplicative factors and the $ \delta $-functions in Eq.~\eqref{eq:22} make it possible to integrate over $ ( p_{i} , q_{i} , r_{i} , s_{i} , t_{i} )$. This leaves us with 
 \begin{equation}\label{eq:25}
     I = \int \mathop{\mathcal{D} \phi_{i}} \mathop{\mathcal{D} \lambda_{i}} \mathop{\mathcal{D} \chi_{i}} \mathop{\mathcal{D} \psi_{i}} \mathop{\mathcal{D} \theta_{i}} \mathop{\mathcal{D} B_{i}} \mathop{\mathcal{D} E_{i}} \mathop{\mathcal{D} G_{i}} \mathop{\mathcal{D} \Xi_{i}} \mathop{\mathcal{D} \Omega_{i}} \exp \left [ i \int \mathop{d x} \left ( \mathcal{L}_{\text{eff}} + \mathcal{L}_{\text{gf}}\right )\right ] \mathop{\rm Sdet} \undertilde{M}_{jp}
 \end{equation}
 with $ \mathcal{L}_{\text{eff}} $ given by Eq.~\eqref{eq:8}, $ \mathcal{L}_{\text{gf}}$ by Eq.~\eqref{eq:20}, and $ \mathop{\rm Sdet} \undertilde{M}_{jp}$ being the Faddeev-Popov superdeterminant of Eq.~\eqref{eq:22}. This functional determinant can be exponentiated using Fermionic ghost fields $ ( \bar{c}_{i} , c_{i} )$, $ ( \bar{d}_{i} , d_{i} )$, $ ( \bar{e}_{i} , e_{i} )$ and Bosonic ghost fields $ ( \tilde{\gamma }_{i} , \gamma_{i} )$, $ ( \tilde{\epsilon }_{i} , \epsilon_{i} )$ \cite{berezin:1987}.\footnote{Note that the complex conjugate of $A_{i}  $ is denoted here by $ \tilde{A}_{i} $.} This leads to 
 \begin{equation}\label{eq:26}
     \begin{split}
         \mathop{\rm Sdet} \undertilde{M}_{jp} ={}& \int 
     \mathop{\mathcal{D} \bar{c}_{i}} \mathop{\mathcal{D} c_{i}}  
     \mathop{\mathcal{D} \bar{d}_{i}} \mathop{\mathcal{D} d_{i}}  
     \mathop{\mathcal{D} \bar{e}_{i}} \mathop{\mathcal{D} e_{i}}  
     \mathop{\mathcal{D} \tilde{\gamma}_{i}} \mathop{\mathcal{D} \gamma_{i}}  
         \mathop{\mathcal{D} \tilde{\epsilon}_{i}} \mathop{\mathcal{D} \epsilon_{i}} \\ &  
     \exp i \int \mathop{d^{}x} \begin{pmatrix}
         \bar{d}_{i} & \bar{c}_{i} & \bar{e}_{i} & \tilde{\gamma}_{i} & \tilde{\epsilon}_{i}
     \end{pmatrix}
     F_{ij}  
\begin{pmatrix}
     0 & H_{jp} & 0 & 0 & 0 \\ 
     H_{jp} & H_{jp} + X_{jp}  & H_{jp,k} \chi_{k}  & -H_{jp,k} \theta_{k} &H_{jp,k} {\psi}_{k}  \\
     0 & H_{jp,k} \chi_{k} & H_{jp} & 0 & 0 \\
     0 & H_{jp,k} \psi_{k} & 0 & H_{jp} & 0 \\
     0 & H_{jp,k} \theta_{k} & 0 & 0 & H_{jp} \\ 
     \end{pmatrix}
\begin{pmatrix}
    d_{p} \\ 
    c_{p} \\
    e_{p} \\
    \gamma_{p} \\
    \epsilon_{p} 
\end{pmatrix},
     \end{split}
 \end{equation}
where we used 
\begin{equation}\label{eq:GF:mLagrange multiplier:supermatrixMprop}
    \mathop{\mathrm{Sdet}}  
\begin{pmatrix}
    0 & A & 0 & 0 & 0 \\ 
    A & B & C  & -D & E  \\
    0 & C & A & 0 & 0 \\
    0 & E  & 0 & A & 0 \\
    0 & D & 0 & 0 & A \\ 
    \end{pmatrix}
=   \mathop{\mathrm{Sdet}}  
\begin{pmatrix}
    0 & A & 0 & 0 & 0 \\ 
    A & A+B & C  & -D & E  \\
    0 & C & A & 0 & 0 \\
    0 & E  & 0 & A & 0 \\
    0 & D & 0 & 0 & A \\ 
    \end{pmatrix}.
\end{equation}
The argument of the exponential in Eq.~\eqref{eq:26} defines the Faddeev-Popov action, $ \int \mathop{dx} \mathcal{L}_{\text{FP}} $.

Note that, 
\begin{equation}\label{eq:ghostdettotal}
\mathop{\mathrm{Sdet}}  
\begin{pmatrix}
    0 & A & 0 & 0 & 0 \\ 
    A & A & C  & -D & E  \\
    0 & C & A & 0 & 0 \\
    0 & E  & 0 & A & 0 \\
    0 & D & 0 & 0 & A \\ 
    \end{pmatrix} 
=   \mathop{\mathrm{det}} A
\end{equation}
(see the appendix C for a derivation). 
Thus, the Faddeev-Popov superdeterminant in the extended Lagrange multiplier formalism is equal to $ \det F_{ij} H_{jp} $, i.e., the usual Faddeev-Popov determinant which is introduced when there is no Lagrange multiplier.

We now will consider the possibility of there being a global gauge invariance (a ``BRST'' invariance \cite{Becchi:1974xu, Tyutin:1975qk}) in the total Lagrangian 
\begin{equation}\label{eq:27}
    \mathcal{L}_{\text{T}} = \mathcal{L}_{\text{eff}} + \mathcal{L}_{\text{gf}} + \mathcal{L}_{\text{FP}} .
\end{equation}

\section{Global gauge invariance}\label{section:GGI}

In keeping with how BRST symmetry is introduced for gauge theories, we begin by replacing the gauge functions $ \xi_{i} $, $ \zeta_{i} $, $ \sigma_{i} $, $ \pi_{i} $ and $ \tau_{i} $ with the ghost fields $ c_{i} $, $ d_{i} $, $ e_{i} $, $ \gamma_{i} $ and $ \epsilon_{i} $ multiplied by $ \eta $, a constant Fermionic scalar, so that the BRST transformations of the classical fields are \begin{subequations}\label{eq:28}
     \begin{align}
         \label{eq:28a}
         &\delta   \phi_{i} = H_{ij} c_{j} \eta , \\
         \label{eq:28b}
         &\delta \lambda_{i} = 
         H_{ij} d_{j} \eta +  \left[ H_{ij,k}\lambda_{k} + H_{ij,mn} \left ( \psi_{m} \theta_{n} + \frac{1}{2} \chi_{m} \chi_{n}\right ) \right]   c_j \eta + H_{ij,k}( \chi_{k} e_{j}   +   \psi_{k} \epsilon_{j}  -  \theta_{k}{\gamma}_{j}) \eta 
         , \\ 
         \label{eq:28c}
         &\delta  \chi_{i} = H_{ij} e_{j} \eta +  H_{ij,k}\chi_{k} c_{j}  \eta , \\
         \label{eq:28d}
         &\delta  \psi_{i} = H_{ij} \gamma_{j} \eta +  H_{ij,k} \psi_{k} c_{j} \eta ,\\
         \label{eq:28e}
         &\delta  \theta_{i} = H_{ij} \epsilon_{j} \eta +  H_{ij,k} {\theta}_{k} c_{j} \eta .
     \end{align}
\end{subequations}
In addition to these BRST transformations, it follows immediately that we also have these transformations: 
\begin{equation}\label{eq:29}
    \delta B_{i} = \delta E_{i} = \delta G_{i} = \delta \Xi_{i} = \delta \Omega_{i} =0
\end{equation}
and 
\begin{equation}\label{eq:30}
    \delta \bar{c}_{i} = - B_{i} \eta, \ \delta \bar{d}_{i} = - E_{i} \eta , \ \delta \bar{e}_{i} = - G_{i} \eta , \ \delta \tilde{\gamma}_{i} = - \Omega_{i} \eta , \ \delta \tilde{\epsilon}_{i} = - \Xi_{i} \eta.
\end{equation}

The transformation of $ c_{i} $, $ d_{i} $, $ e_{i} $, $ \gamma_{i} $ and $ \delta_{i} $ now are determined by the requirements that \begin{subequations}\label{eq:31}
    \begin{align} 
        \label{eq:31a}
        \delta ( H_{ij} c_{j} ) ={}& 0 \\
        \label{eq:31b}
        \delta \left[H_{ij} d_{j}  +  \left[ H_{ij,k}\lambda_{k} + H_{ij,mn} \left ( \psi_{m} \theta_{n} + \frac{1}{2} \chi_{m} \chi_{n}\right ) \right]   c_j  + H_{ij,k}( \chi_{k} e_{j}   +   \psi_{k} \epsilon_{j}  -  \theta_{k}{\gamma}_{j})  \right] ={}&0 \\
        \label{eq:31c}
        \delta ( H_{ij} e_{j}  +  H_{ij,k}\chi_{k} c_{j}  ) ={}&0 , \\
        \label{eq:31d}
        \delta (H_{ij} \gamma_{j}  +  H_{ij,k} \psi_{k} c_{j}  ) ={}&0,\\
        \label{eq:31e}
        \delta ( H_{ij} \epsilon_{j}  +  H_{ij,k} {\theta}_{k} c_{j}  ) ={}&0 .
    \end{align}
\end{subequations}

In order to do this, we must first impose conditions on $ H_{ik} ( \phi_{i} )$. We consider the commutator of two gauge transformations of the form of Eq.~\eqref{eq:9}, 
\begin{equation}\label{eq:32}
    [ \delta_{\lambda} , \delta_{\sigma} ] \phi_{i} = \left [\left( \frac{\partial H_{ij}}{\partial \phi_{k}} \sigma_{j} \right) ( H_{kl} \lambda_{l} ) 
    - \left ( \frac{\partial H_{ij}}{\partial \phi_{k}} \lambda_{j}\right ) H_{kl} \sigma_{l}\right ],
\end{equation}
which is itself a gauge transformation, so that 
\begin{equation}\label{eq:33}
    \left ( H_{ij , k} H_{kl} - H_{il , k} H_{kj}\right ) \sigma_{j} \lambda_{l} = f_{jl|k} H_{ik} \lambda_{l} \sigma_{j}
\end{equation}
if the gauge transformation is ``closed''. For an ``open'' gauge transformation, Eq.~\eqref{eq:33} is satisfied only if $ \phi_{i} $ satisfies the classical equation of motion (``on the mass shell'').

We will only consider gauge transformations that are closed with the additional restriction that $ H_{ij} ( \phi_{i} )$ is at most linear in $ \phi_{i} $ so that \begin{subequations}\label{eq:34}
    \begin{align}
        \label{eq:34a}
        H_{ij , m n} ={}&0, \\
        \label{eq:34b}
        f_{jl|k, i} ={}&0.
    \end{align}
\end{subequations}
Gauge theories that satisfy these conditions are the so-called YM type theories \cite{Lavrov:2019nuz}. Besides the YM theory, the EH action is another interesting example of a gauge theory of this type. In the next section, we will consider the quantization of the YM in the extended Lagrange multiplier formalism. 

From Eq.~\eqref{eq:31a}, we obtain that 
\begin{equation}\label{eq:35}
    \begin{split}
        \delta ( H_{ij} c_{j} ) ={}& \frac{\partial H_{ij}}{\partial \phi_{k}} ( H_{kl} c_{l} \eta ) c_{j} + H_{ij} \delta c_{j} \\
        ={}& - \frac{1}{2} \left ( \frac{\partial H_{ij}}{\partial \phi_{k}} H_{kl} - \frac{\partial H_{il}}{\partial \phi_{k}} H_{kj}\right ) c_{l} c_{j} \eta + H_{ij} \delta c_{j} .
    \end{split}
\end{equation}
From Eqs.~\eqref{eq:33} and Eq.~\eqref{eq:35}, we see that 
\begin{equation}\label{eq:36}
    \delta c_{j} = - \frac{1}{2} f_{mn | j} c_{m} c_{n} \eta.
\end{equation}

To obtain $ \delta e_{j} $, we now examine Eq.~\eqref{eq:31c}, 
\begin{equation}\label{eq:37}
    H_{ij,k} \delta \phi_{l} \chi_{k} c_{j} + H_{ij,k} ( \delta \chi_{k} c_{j} + \chi_{k} \delta c_{j} ) + H_{ij, k} \delta \phi_{k} e_{j} + H_{ij} \delta e_{j} =0. 
\end{equation}
Upon using Eqs.~\eqref{eq:34a}, \eqref{eq:28c} and \eqref{eq:36}, \eqref{eq:37} becomes 
\begin{equation}\label{eq:38}
    H_{ij,k} \left [ \left ( \frac{\partial H_{kl}}{\partial \phi_{m}} \chi_{m} c_{l} + H_{kl} e_{l}\right ) \eta c_{j} + \chi_{k} \left ( - \frac{1}{2} f_{m n|j} c_{m} c_{n} \eta\right )\right ] + H_{ij,k} ( H_{kl} c_{l} \eta ) e_{j} + H_{ij} \delta e_{j} =0.
\end{equation}
Eq.~\eqref{eq:33} reduces \eqref{eq:38} to simply \begin{subequations}\label{eq:39}
    \begin{align} \label{eq:39a}
        \delta e_{j} ={}& - f_{m n | j} c_{m} e_{n} \eta. \\ \intertext{In a similar way, Eqs.~\eqref{eq:31d} and \eqref{eq:31e} lead to}
        \label{eq:39b}
        \delta \gamma_{j} ={}&- f_{mn |j} c_{m} \gamma_{n} \eta , \\
        \label{eq:39c}
        \delta \epsilon_{j} ={}&- f_{mn |j} c_{m} \epsilon_{n} \eta .\\
        \intertext{Finally, from Eq.~\eqref{eq:31b}, we obtain}
        \label{eq:39d}
        \delta d_{j} ={}&- f_{mn |j} \left ( d_{m} c_{n} + \frac{1}{2} e_{m} e_{n} + \gamma_{m} \epsilon_{n}\right ) \eta .
    \end{align}
\end{subequations}

In the appendix D, we find the corresponding Zinn-Justin master equation \cite{Zinn-Justin:1974ggz} which follows from the global gauge invariance in Eqs.~\eqref{eq:28}, \eqref{eq:29}, \eqref{eq:30}, \eqref{eq:36} and \eqref{eq:39}. 

\section{Yang-Mills theory}\label{section:YMt}

The classical YM Lagrangian reads
\begin{equation}\label{eq:61}
    \mathfrak{L}_{\text{cl}}= - \frac{1}{4} F_{\mu \nu}^{a} F^{a \, \mu \nu},
\end{equation}
where 
\begin{equation}\label{eq:62}
F_{\mu \nu}^{a} = \partial_{\mu} A_{\nu}^{a} - \partial_{\nu} A_{\mu}^{a} + g f^{abc} A_{\mu}^{b} A_{\nu}^{c} .
\end{equation}
It is invariant under the gauge transformation 
\begin{equation}\label{eq:63}
    A_{\mu}^{a} \to A_{\mu}^{' a} = A_{\mu}^{a} + D_{\mu}^{ab} (A) \xi^{b}
\end{equation}
in which $ D_{\mu}^{ab} (A) \equiv \partial_{\mu} \delta^{ab} + g f^{apb} A_{\mu}^{p} $. Comparing Eq.~\eqref{eq:63} with Eq.~\eqref{eq:9}, we can identify $ H_{ij} ( \phi_{i})$ with $ D_{\mu}^{ab} (A)$. Since $ D_{\mu}^{ab} (A)$ satisfies the conditions \eqref{eq:34} and we restrict the gauge group to compact semisimple Lie groups, the quantization of the YM theory in the extended Lagrange multiplier formalism follows identically.


Replacing the YM classical Lagrangian in Eq.~\eqref{eq:61} into Eq.~\eqref{eq:8} yields the classical effective YM Lagrangian in the framework of the extended Lagrange multiplier formalism:
\begin{equation}\label{eq:64}
    \begin{split}
        \mathfrak{L}_{\text{eff}} ={}&\mathfrak{L}_{\text{cl}}+ \lambda_{\mu}^{a} D^{ab }_{\nu}  F^{ b \nu \mu}  - \frac{1}{4}  ( \partial_{\mu} \chi_{\nu}^{a}  - \partial_{\nu} \chi_{\mu}^{a} )^{2} 
- \frac{1}{2}  
( \partial_{\mu} \psi_{\nu}^{a}  - \partial_{\nu} \psi_{\mu}^{a} ) 
( \partial^{\mu} \theta^{ a \, \nu }  - \partial^{\nu} \theta^{a \, \mu } ) \\ 
                                     &  
                                     - g f^{abc} ( \partial_{\mu} \chi_{\nu}^{a}  - \partial_{\nu} \chi_{\mu}^{a} ) A^{b \, \mu} \chi^{c \, \nu} 
                                     - \frac{g}{2} f^{abc}  (\partial_{\mu} A_{\nu}^{a} - \partial_{\nu} A_{\mu}^{a} )    \chi^{b \, \mu} \chi^{c \, \nu} 
                                  \\ & 
                                  - \frac{g^{2}}{2}   f^{abc} f^{ade} A_{\mu}^{b} \chi_{\nu}^{c} A^{d \, \mu} \chi^{e \, \nu}
                                  - \frac{g^{2}}{2}   f^{abc} f^{ade} A_{\mu}^{b} \chi_{\nu}^{c} \chi^{d \, \mu} A^{e \, \nu}
                                    - \frac{g^{2}}{2} f^{abc} f^{ade} A_{\mu}^{b} A_{\nu}^{c} \chi^{d \, \mu} \chi^{e \, \nu} 
                                     \\ & 
- g f^{abc} ( \partial_{\mu} \psi_{\nu}^{a}  - \partial_{\nu} \psi_{\mu}^{a} ) A^{b \, \mu} \theta^{c \, \nu} 
- g f^{abc}  A^{b \, \mu} \psi^{c \, \nu}  ( \partial_{\mu} \theta_{\nu}^{a}  - \partial_{\nu} \theta_{\mu}^{a} )
- g  f^{abc}  (\partial_{\mu} A_{\nu}^{a} - \partial_{\nu} A_{\mu}^{a} )    \psi^{b \, \mu} \theta^{c \, \nu} 
                                  \\ & 
                                  - g^2   f^{abc} f^{ade} A_{\mu}^{b} \psi_{\nu}^{c} A^{d \, \mu} \theta^{e \, \nu}
                                  - g^2   f^{abc} f^{ade} A_{\mu}^{b} \psi_{\nu}^{c} \theta^{d \, \mu} A^{e \, \nu}
                                  - g^2 f^{abc} f^{ade} A_{\mu}^{b} A_{\nu}^{c} \psi^{d \, \mu} \theta^{e \, \nu}.
    \end{split}
\end{equation}
By Eqs.~\eqref{eq:16} and \eqref{eq:17}, the gauge invariance \eqref{eq:63} is now accompanied by 
 \begin{subequations}\label{eq:66}
 \begin{align}
     \lambda_{\mu}^a \to \lambda_{\mu}^{'a} ={}& \lambda_{\mu}^a +  
     g f^{abc} \lambda^{b} \xi^c  , 
 \\
 \psi_{\mu}^a \to \psi_{\mu}^{'a} ={}& \psi_{\mu}^a + g f^{abc} \psi^{b}  \xi^c, \\
 \theta_{\mu}^a \to \theta_{\mu}^{'a} ={}& \theta_{\mu}^a  
 + g f^{ab c} \theta^{b}  \xi^c,\\
 \chi_{\mu}^a \to \chi_{\mu}^{'a}  ={}& \chi_{\mu}^a 
 + g f^{abc}  \chi^{b} \xi^c,
     \end{align} 
 \end{subequations}
and 
\begin{subequations}\label{eq:67}
    \begin{align}
        \lambda_{\mu}^a \to \lambda_{\mu}^{'a} ={}&\lambda_{\mu}^a  + D_{\mu}^{ab}(A)  \zeta^{b}+ g f^{apb} \left ( \chi^{p}_{\mu}   \sigma^{b}  + \psi^{p}_{\mu}   \tau^{b}  - \theta^{p}_{\mu}   \pi^{b}  \right ), \\ 
        \psi_{\mu}^a \to \psi_{\mu}^{'a} ={}& \psi_{\mu}^a + D_{\mu}^{ab}(A) \pi^b, \\
        \theta_{\mu}^a \to \theta_{\mu}^{'a} ={}&\theta_{\mu}^a + D_{\mu}^{ab}(A) \tau^b ,\\
        \chi_{\mu}^a \to \chi_{\mu}^{'a} ={}& \chi_{\mu}^a + D_{\mu}^{ab}(A) \sigma^b .
     \end{align}
 \end{subequations}

 Now, in order to quantize the action in Eq.~\eqref{eq:64}, we will employ the gauge fixing 
 \begin{equation}\label{eq:68}
     F^{\mu ab} A_{\mu}^{b} \equiv \partial^{\mu} A_{\mu}^{a} = \partial^{\mu} \lambda_{\mu}^{a} = \partial^{\mu} \chi_{\mu}^{a} = \partial^{\mu} \psi_{\mu}^{a} = \partial^{\mu} \theta_{\mu}^{a} =0
 \end{equation}
 leading to the gauge fixing Lagrangian (see Eq.~\eqref{eq:19})
\begin{equation}\label{eq:69}
    \mathfrak{L}_{\text{gf}} = - \frac{1}{2 \alpha} (\partial \cdot A^{a} )^{2} - \frac{1}{\alpha} (\partial \cdot \lambda^{a} ) (\partial \cdot A^{a} ) - \frac{1}{2 \alpha } (\partial \cdot \chi^{a} )^{2} - \frac{1}{\alpha} (\partial \cdot \psi^{a} )(\partial \cdot \theta^{a} ),
\end{equation}
where $ \partial \cdot X_{I} \equiv \partial_{\mu} X^{\mu}_{I} $ ($I$ are internal indices).  

In the YM theory, we identify $ H_{ij} ( \phi ) \mapsto  D_{\mu}^{ab} (A)$ and $ F_{ij} \mapsto F^{\mu ab} $. Using these relation in Eq.~\eqref{eq:22}, we find that the Faddeev-Popov superdeterminant is 
\begin{equation}\label{eq:GF:YM1FP}
    \mathop{\mathrm{Sdet}}  \underaccent{\tilde}{M}_{jk} = \begin{pmatrix}
        0 & \partial \cdot D^{ab} (A) & 0 & 0 & 0 \\ 
        \partial \cdot D^{ab} (A)  & \partial \cdot D^{ab} (A + \lambda ) & g f^{apb} \partial \cdot \chi^p  & -g f^{apb} \partial \cdot \theta^p& g f^{apb} \partial \cdot {\psi}^p  \\
    0 & g f^{apb} \partial \cdot \chi^p &\partial \cdot D^{ab}(A)& 0 & 0 \\
    0 & g f^{apb} \partial \cdot {\psi}^p  & 0 &\partial \cdot D^{ab}(A)& 0 \\
    0 & g f^{apb} \partial \cdot \theta^p& 0 & 0 &\partial \cdot D^{ab}(A)\\ 
    \end{pmatrix},
\end{equation}
where we used that $ H_{ij,k} = g f^{ikj} $. 
We obtain the ghost Lagrangian by replacing Eq.~\eqref{eq:GF:YM1FP} in Eq.~\eqref{eq:26} which reads 
\begin{equation}\label{eq:610}
    \begin{split}
        \mathfrak{L}_{\text{gh}} ={}&\bar{c}^{a} \partial \cdot D^{ab} (A + \lambda ) c^{b} + \bar{d}^{a} \partial \cdot D^{ab} (A) c^{b} + \bar{c}^{a} \partial \cdot D^{ab} (A) d^{b} 
    \\ &  \quad \quad \quad  + \bar{e}^{a} \partial \cdot D^{ab} (A)e^{b} + \tilde{\gamma}^{a} \partial \cdot D^{ab} (A)\gamma^{b} + \tilde{\epsilon}^{a} \partial \cdot D^{ab} (A)\epsilon^{b} \\ &  
        \quad \quad \quad + \bar{c}^{a} g f^{apb} \partial \cdot \chi^p e^{b} + \bar{e}^{a} g f^{apb} \partial \cdot \chi^p c^{b}    
        -\bar{c}^{a}  g f^{apb} \partial \cdot \theta^p \gamma^{b}\\ &  \quad \quad \quad + \tilde{\gamma}^{a} g f^{apb} \partial \cdot {\psi}^p c^{b}  
                                +\bar{c}^{a} g f^{apb} \partial \cdot {\psi}^p \epsilon^{b} + \tilde{\epsilon}^{a} g f^{apb} \partial \cdot \theta^p c^{b}.
\end{split}
\end{equation}

Thus, the generating functional of the YM theory in this framework is given by
\begin{equation}\label{eq:611}
    Z [ \bm{J} , \bm{\eta} , \bar{\bm{\eta}} ]   = \int \mathop{\mathcal{D} \bm{A}_{\mu}^{a}} \mathop{\mathcal{D} \bm{c}^{a} } \mathop{\mathcal{D} \bar{\bm{c}}^{a} } \exp i \int \mathop{d x} \left ( \mathfrak{L}_{ \text{eff} } + \mathfrak{L}_{\text{gf} } + \mathfrak{L}_{\text{gh}} + \bar{\bm{J}}^{a \, \mu} \bm{A}_{\mu}^{a} + \bar{\bm{\eta}}^{a} \bm{c}^{a} +  \bar{\bm{c}}^{a} \bm{\eta}^{a}  \right ), 
\end{equation}
where we have used a compact notation in which $ \bm{A}_{\mu}^{a} = ( A_{\mu}^{a} , \lambda_{\mu}^{a} , \chi_{\mu}^{a} , \psi_{\mu}^{a} , \theta_{\mu}^{a} )$, $ \bm{c}^{a} = ( c^{a} , d^{a} , e^{a} , \gamma^{a} , \epsilon^{a} )$ and 
\begin{equation}\label{eq:612}
    \mathop{\mathcal{D} \bm{A}_{\mu}^{a} } \equiv\mathop{\mathcal{D} A_{\mu}^{a} } \mathop{\mathcal{D} \lambda_{\mu}^{a}} \mathop{\mathcal{D} \chi_{\mu}^{a}} \mathop{\mathcal{D} \psi_{\mu}^{a}} \mathop{\mathcal{D} \theta_{\mu}^{a}}, \quad 
    \mathop{\mathcal{D} \bm{c}^{a} } \equiv 
    \mathop{\mathcal{D} c^{a}}     
    \mathop{\mathcal{D} d^{a}}          
    \mathop{\mathcal{D} e^{a}}            
    \mathop{\mathcal{D} \gamma^{a}}   
    \mathop{\mathcal{D} \epsilon^{a}},  \quad 
    \mathop{\mathcal{D} \bar{\bm{c}}^{a} } \equiv 
    \mathop{\mathcal{D} \bar{c}^{a}} 
    \mathop{\mathcal{D} \bar{d}^{a}} 
    \mathop{\mathcal{D} \bar{e}^{a}} 
    \mathop{\mathcal{D} \tilde{\gamma}^{a}} 
    \mathop{\mathcal{D} \tilde{\epsilon}^{a}}.
\end{equation}
The sources terms are defined as
\begin{subequations}\label{eq:613}
    \begin{align}
        \bar{\bm{J}}^{a \, \mu} \bm{A}_{\mu}^{a} \equiv{}& J^{a \, \mu} (A_{\mu}^{a} + \lambda_{\mu}^{a} )+ K^{a \, \mu} \lambda_{\mu}^{a} + L^{a \, \mu} \chi_{\mu}^{a} + \bar{\eta}^{a \, \mu} \psi_{\mu}^{a} + \bar{\kappa}^{a \, \mu} \theta_{\mu}^{a}, \\
        \bar{\bm{\eta}}^{a} \bm{c}^{a} \equiv{}& \bar{\eta}^{a} (c^{a} + d^{a} )+ \bar{\kappa}^{a} d^{a}  + \bar{\upsilon}^{a} e^{a}   + \tilde{J}^{a} \gamma^{a} + \tilde{K}^{a} \epsilon^{a}, \\
        \bar{\bm{c}}^{a} \bm{\eta}^{a} \equiv{}& ( \bar{c}^{a} + \bar{d}^{a} ) \eta^{a} + \bar{d}^{a} \kappa^{a} + \bar{e}^{a} \upsilon^{a} + \tilde{\gamma}^{a} J^{a} + \tilde{\epsilon}^{a} K^{a},
\end{align}
\end{subequations}
where ($ J^{a \, \mu} $, $ K^{a \, \mu} $, $ L^{a \, \mu} $) are real ordinary sources, ($ {J}^{a} $, $ {K}^{a} $) are complex ordinary sources, and ($ \eta^{a \, \mu} $, $ \kappa^{a \, \mu} $, $ \eta^{a} $, $ \kappa^{a} $, $ \upsilon^{a} $) are Fermionic sources. The Feynman rules derived from Eq.~\eqref{eq:611} are presented in appendix B.

\subsection{BRST transformation}\label{section:YMBRST}

We also can obtain the BRST transformation that leaves the action in Eq.~\eqref{eq:611} invariant. 
Replacing $H_{ij} ( \phi_{i}) \mapsto D_{\mu}^{ab} (A) $ in Eqs.~\eqref{eq:28}, \eqref{eq:33}, \eqref{eq:36} and \eqref{eq:39}, we find that the total Lagrangian in Eq.~\eqref{eq:611} must be invariant under the following transformation\footnote{In the appendix B, the gauge algebra of the extended Lagrange multiplier theory is studied. The gauge algebra of the extended Lagrange multiplier theory is closed when the gauge algebra of the starting theory is closed and the conditions \eqref{eq:34} are satisfied.} 
    \begin{subequations}\label{eq:BRSTYM}
     \begin{align}
         \delta   A_{\mu}^{a}  ={}& D_{\mu}^{ab}(A) c^{b}   \eta , \\
         \delta \lambda_{\mu}^{a} ={}& 
         D_{\mu}^{ab} (A)d^{b}  \eta 
       +   g f^{abc} \lambda^{b} c^{c}  \eta 
       + g f^{abc} ( \chi^b e^c   +   \psi^b \epsilon^c  -  \theta^b{\gamma}^c) \eta 
         , \\ 
         \delta  \chi_{\mu}^a ={}& D_{\mu}^{ab} (A)e^b \eta +  g f^{abc}\chi^b c^c  \eta , \\
         \delta  \psi_{\mu}^a ={}& D_{\mu}^{ab}(A) \gamma^b \eta +  g f^{abc} \psi^b c^c \eta ,\\
         \delta  \theta_{\mu}^a ={}&D_{\mu}^{ab} (A)\epsilon^b \eta +  g f^{abc} {\theta}^b c^c \eta,
        \\ 
         \delta c_{\mu}^a ={}&  -\frac{g}{2} f^{abc}  c^b c^c \eta\\        
        \delta e_{\mu}^a ={}& - g f^{abc}  c^b e^c \eta \\
        \delta \gamma_{\mu}^a ={}&- g f^{abc} c^b \gamma^c \eta , \\
        \delta \epsilon_{\mu}^a ={}&- gf^{abc} c^b \epsilon^c \eta \\
        \delta d_{\mu}^a ={}&- gf^{abc} \left ( d^b c^c + \frac{1}{2} e^b e^c + \gamma^b \epsilon^c\right ) \eta,
     \end{align}
\end{subequations}
and we also have 
\begin{equation}\label{eq:614}
    \delta \bar{c}^a =- \frac{1}{\alpha} \partial \cdot {(A + \lambda )}^{a}  \eta, \ \delta \bar{d}^a = -\frac{1}{\alpha}\partial \cdot A^{a} \eta , \ \delta \bar{e}^a = -\frac{1}{\alpha}\partial \cdot \chi^{a}  \eta , \ \delta \tilde{\gamma}^a =  -\frac{1}{\alpha} \partial \cdot \theta^{a}  \eta , \ \delta \tilde{\epsilon}^a =  \frac{1}{\alpha} \partial \cdot \psi^a \eta.
\end{equation}
This transformation can be used to show that unitarity is retained in the YM theory in the framework of the extended Lagrange multiplier theory~\cite{Brandt:2021qgh}. 

\section{Discussion}\label{section:discussion}

It has been shown \cite{McKeon:1992rq, Brandt:2018lbe, Brandt:2019ymg} that by using a Lagrange multiplier field to ensure that the equations of motion are satisfied, radiative effects beyond one-loop order are suppressed. Furthermore, it has been demonstrated that the path integral associated with this action becomes form invariant under a change of variable if the measure of the path integral is suitably altered. It follows that the resulting effective action coincides exactly with the usual one-loop effective action \cite{Brandt:2022kjo}. In this paper, we have extended this result to theories in which the classical action possesses a gauge invariance. The gauge transformations in the tree level action (including all Lagrange multiplier and ghost fields) as well as the BRST transformation have been derived.

Having the quantum effects beyond one-loop order suppressed is of particular relevance when discussing quantum gravity. In quantum electrodynamics and YM theory, the divergences that arise due to quantum effects are proportional to terms appearing in the initial action, and hence they can be absorbed into these terms through the process of ``renormalization'' \cite{Dyson:1949bp, tHooft:1971akt, tHooft:1971qjg}. When the EH action is quantized using the Faddeev-Popov procedure, then at one-loop order divergences have a different character; they vanish when external fields satisfy the classical equation of motion (are ``on mass-shell'') \cite{tHooft:1974toh}. This no longer holds at two-loop order and beyond \cite{Goroff:1985th, vandeVen:1991gw}. However, if the EH action is supplemented by a Lagrange multiplier term, then this one-loop divergence can be absorbed by the Lagrange multiplier field \cite{Brandt:2018lbe}. No difficulty arises at higher loop order as no effects beyond one-loop order arise. The resulting effective action is free of divergences, and as it possesses a BRST invariance, it is also unitary. This approach can also be employed when matter fields are present in addition to the gravitational field \cite{Brandt:2021qgh}.

We note that when a Lagrange multiplier field is used in a similar way with a YM theory, the divergences which now arise exclusively at one-loop order, lead to closed-form expressions for the renormalization group $ \beta $-function associated with the running coupling. Since at one-loop order, this running coupling develops unphysical ``Landau poles'', one cannot restrict YM theory to one-loop order using a Lagrange multiplier field in a physically consistent manner. This difficulty does not arise when one uses a Lagrange multiplier field in conjunction with EH action. 

However, YM and gravity (based on the EH action) are both non-linear gauge theories, which have several other well-known similarities \cite{Bern:2010ue,  Monteiro:2014cda,White:2017mwc}. With this in mind,  we have considered the YM theory in the extended Lagrange multiplier theory to illustrate the general procedure introduced in Section III. 
With the usual covariant gauge fixing condition \eqref{eq:68}, which is used to fix all the gauge invariances present in the classical effective action in Eq.~\eqref{eq:64}, we have obtained the generating functional of the YM theory in the framework of the extended Lagrange multiplier theory \eqref{eq:611}. We also found the BRST transformation in Eqs.~\eqref{eq:BRSTYM} and \eqref{eq:614} that leaves the
resulting quantum effective action $ \mathfrak{L}_{\text{eff}} + \mathfrak{L}_\text{gf} + \mathfrak{L}_{ \text{gh} } $, which appears in Eq.~\eqref{eq:611}, invariant. 

The principle result of this paper is that when a Lagrange multiplier field is used to eliminate quantum effects beyond one-loop order in a gauge theory, and a suitable modification of the measure of the path-integral is made, then the path integral is form-invariant under a change of integration variables, that the effective action possesses a BRST invariance if the gauge algebra is closed, and that the quantum effects coincide with the usual one-loop effects arising from the classical action alone. This is of special significance when using the Faddeev-Popov procedure to quantize the EH action (when using the extended Lagrange multiplier theory) as it results in a theory that is unitary and renormalizable with the classical EH action intact. There is no need to postulate the existence of extra dimensions or degrees of freedom, or to invoke special non-perturbative effects (see, for example, \cite{Nagy:2012ef}).

We note that the approach outlined in Section III can be straightforwardly applied to the case of the EH action. The EH action is invariant under diffeomorphisms in which the metric transforms as
\begin{equation}\label{eq:VI1}
    g^{\mu \nu} \to (g^{\mu \nu})^{\prime}   = (- \partial_{\lambda}g^{\mu \nu} 
    + g^{\alpha \nu} \delta_{\lambda}^{\mu}  \partial_{\alpha} 
    + g^{\mu \alpha } \delta_{\lambda}^{\nu}  \partial_{\alpha}) \xi^{\lambda}\equiv  \tensor{H}{^{ \mu \nu }_{\lambda} } ( g^{\mu \nu} ) \xi^{\lambda} ,
\end{equation}
which enables us to identify $ g^{\mu \nu} $ and $\tensor{H}{^{\mu \nu}_{\lambda}} ( g^{\mu \nu} ) $ in Eq.~\eqref{eq:VI1} with
$\phi_{i} $ and $ H_{ij} ( \phi_{i} )$ in Eq.~\eqref{eq:9}. 

However, in the case of the Einstein-Cartan action \cite{Kibble:1961ba, Sciama:1964jqa}, which describes spacetimes with torsion, applying this approach is non-trivial.  
This action, which is described in terms of the tetrad field $ e_{\mu}^{a} $ and the spin connection $ \omega_{\mu ab} $, is invariant under two distinct gauge transformations, namely, the diffeomorphisms and local Lorentz transformations \cite{Hehl:1976kj, Shapiro:2001rz}. 
Hence, the quantization of this theory, in the extended Lagrange multiplier formalism, requires an extension of the Faddeev-Popov procedure outlined in Section III to accommodate two gauge symmetries. This extension would be analogous to what was done in Section III of \cite{Brandt:2024rsy} for the first order form of the Einstein-Cartan theory.\footnote{In Ref. \cite{Brandt:2024rsy} it also has been shown that the Einstein-Cartan theory in first order form is a YM type theory. Hence, the conditions \eqref{eq:34} are satisfied in this theory.}

\begin{acknowledgments}
We thank James P. Edwards and Carmelo P. Martin for useful correspondence.
F.\ T.\ Brandt and S.\ M.-F.\ thank CNPq (Brazil) for financial support. 
\end{acknowledgments}

\appendix

\section{Invariance under field redefinitions}\label{section:IUFR}

To illustrate invariance under field redefinitions, let us consider \begin{align*} 
        I ={}& \int \frac{dx}{2 \pi} \frac{d \lambda}{2 \pi} [ f''(x) ]^{1/2} \exp i \left( f(x) + \lambda f'(x)\right) 
        \\
        ={}& \int \frac{dx}{2 \pi} [ f''(x) ]^{1/2} \delta ( f'(x))\exp i f(x). 
\end{align*}
If $ y(x)$ is a monotonically increasing function, then under the field redefinition $ x \to x' =y(x) $: \[ I = 
    \int \frac{dy}{2 \pi \frac{dy}{dx}} \frac{\delta \left( \frac{df(y) }{dy} \right)}{ \frac{dy}{dx} }\left[ \left( \frac{d y}{d x} \frac{d}{dy}  \right)^{2} f(y) \right]^{1/2} \exp i f(y).
\]  
Since $ f'(y) \delta (f'(y)) = 0$, this becomes \[
    I=\int \frac{dy}{2 \pi \frac{dy }{dx}} \delta ( f'(y)) ( f''(y))^{1/2} \exp i f (y)
\]
as expected. 

\section{Perturbative aspects of the Lagrange multiplier formalism}\label{section:PAofLM}

In this appendix we review how a Lagrange multiplier field can be used to eliminate loop effects beyond one-loop order, as has been discussed in Refs. \cite{Brandt:2018lbe, Brandt:2019ymg, McKeon:1992rq,Brandt:2021qgh, Brandt:2020gms, Brandt:2021nev}.

If we initially consider the action 
\begin{equation}\label{eq:appB1}
    S = \int \mathop{d x} \mathcal{L} ( \phi_{i} )
\end{equation}
for a field $ \phi_{i} $, without gauge symmetries, we modify this by adding a term in which a Lagrange multiplier field $ \lambda_{i} $ is used to impose the classical equation of motion 
\begin{equation}\label{eq:appB2}
    S_{\lambda} = \int \mathop{d x} \left ( \mathcal{L} ( \phi_{i} ) + \lambda_{i} \mathcal{L}_{,i} ( \phi_{i} )\right ) .
\end{equation}
A formal argument shows that if $ \phi_{i} $ and $ \lambda_{i} $ are both treated as quantum fields, then radiative effects beyond one-loop order do not arise. One considers the generating functional 
\begin{equation}\label{eq:appB3}
    Z [ j_{i} ] = \int \mathop{\mathcal{D} \phi_{i}} \mathop{\mathcal{D} \lambda_{i}} \exp{i \int \mathop{d x} \left ( \mathcal{L} ( \phi_{i} ) + \lambda_{i} \mathcal{L}_{, i} ( \phi_{i} ) + j_{i} \phi_{i}\right )}
\end{equation}
and first integrates over $ \lambda_{i} $ to obtain 
\begin{equation}\label{eq:appB4}
    Z [ j_{i} ] = \int \mathop{\mathcal{D} \phi_{i}} \mathop{\mathcal{D} \lambda_{i}} \mathop{\delta} \left (  \mathcal{L}_{, i} ( \phi_{i} )\right ) \exp{i \int \mathop{d x} \left ( \mathcal{L} ( \phi_{i} )  + j_{i} \phi_{i}\right )}.
\end{equation}
A functional extension of the standard formula 
\begin{equation}\label{eq:appB5}
    \int_{- \infty}^{ \infty} \mathop{d x} \mathop{\delta} ( f(x)) g (x) = \sum_{i}^{} g ( \bar{x}_{i} ) | f'( \bar{x}_{i} ) |^{-1},
\end{equation}
where $ f ( \bar{x}_{i} ) = 0$ can be now used with Eq.~\eqref{eq:appB4} to obtain 
\begin{equation}\label{eq:appB6}
    Z[j_{i}] = \sum_{a}^{} \exp i \int \mathop{d x} \left ( \mathcal{L} ( \bar{\phi}_{i}^{a} ) + j_{i} \bar{\phi}_{i}^{a}\right ) |\det \mathcal{L}_{, ij} ( \bar{\phi}_{i}^{a} )|^{-1},
\end{equation}
where $ \mathcal{L}_{, i} ( \bar{\phi}_{i}^{a} ) =0$ defines $ \bar{\phi}_{i}^{a} $. In Eq.~\eqref{eq:appB6}, the exponential is the sum of all tree diagrams while the functional determinant is the square of all one-loop diagrams. No further contributions which would correspond to higher loop contributions to $Z$ arise.

\begin{figure}[ht]
    \includegraphics[scale=0.9]{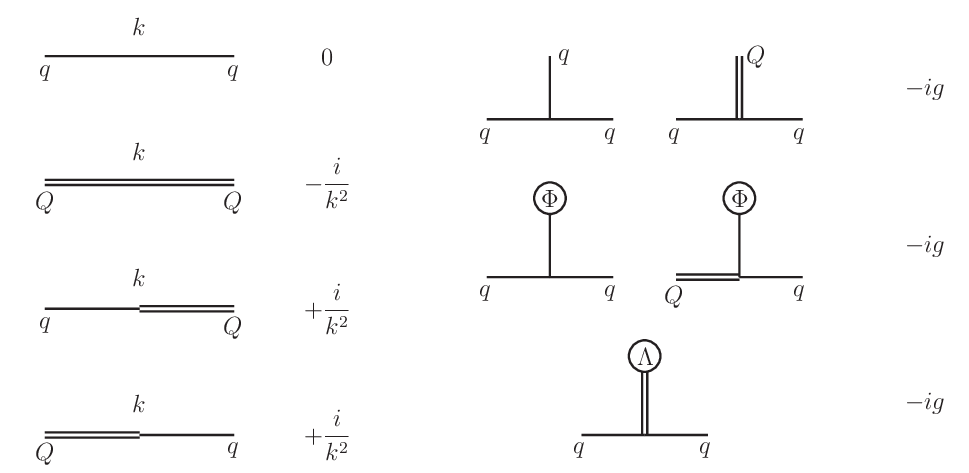}
    \caption{Feynman rules of the scalar model.}\label{fig:FRscalar}
\end{figure}

In order to illustrate perturbatively how the Lagrange multiplier field serves to eliminate higher loop corrections, let us consider a simple scalar model 
\begin{equation}\label{eq:appB7}
    \mathcal{L} ( \phi , \lambda ) = \frac{1}{2} ( \partial_{\mu} \phi )^{2} - \frac{g}{3!} \phi^{3} + \lambda \left( - \partial^{2} \phi - \frac{g}{2!} \phi^{2 }\right)
\end{equation}
and apply background field quantization \cite{Abbott:1980hw, abbott:1982}. This involves considering a source $k$ for the field $ \lambda $, in addition to $j$, the source for $ \phi $, so that (with $ \hbar $ being restored) 
\begin{equation}\label{eq:appB8}
    \begin{split}
        Z [j,k] ={}&\int \mathop{\mathcal{D} \phi} \mathop{\mathcal{D} \lambda} \exp{\frac{i}{\hbar} \int \mathop{d x} \left ( \mathcal{L}(\phi , \lambda ) + j \phi +  k \lambda\right ) } \\
        ={}& \exp \frac{i}{\hbar} W[j,k].
    \end{split}
\end{equation}
Upon performing the usual Legendre transform on $ W [j,k]$, we obtain $ \Gamma [ \Phi , \Lambda ]$ where $ \Phi $ and $ \Lambda $ are background fields for $ \phi $ and $ \lambda $, ($ \phi = \Phi + q$, $ \lambda = \Lambda + Q$) 
\begin{equation}\label{eq:appB9}
    \exp \frac{i}{\hbar} \Gamma [ \Phi , \Lambda ] = \int \mathop{\mathcal{D} q } \mathop{\mathcal{D} Q} \exp \frac{i}{\hbar} \int \mathop{d x} \left [ \mathcal{L} ( \Phi + q, \Lambda + Q) + j q + k Q\right ].
\end{equation}
With $ \mathcal{L} $ given by Eq.~\eqref{eq:appB7}, we find that the Feynman rules that follow from Eq.~\eqref{eq:appB9} are in Fig.~\ref{fig:FRscalar}.
Since there is no propagator for $q$, and since $ \lambda $ and $ Q$ only enter linearly in any vertex, it is impossible to draw a Feynman diagram with more than one loop or with external fields $ \lambda $. The only two and three point functions are in fig. 2.
\begin{figure}[ht]
    \includegraphics[scale=0.8]{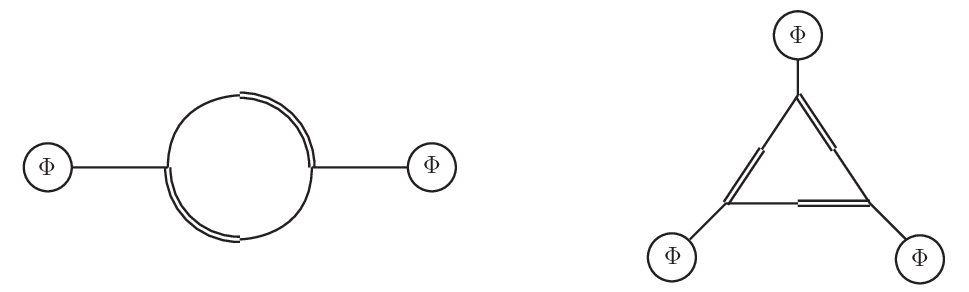}
    \caption{The two- and tree-point functions one-loop contributions.}\label{fig:23scalar}
\end{figure}
All $N$-point one-loop diagrams can easily be constructed; their sum yields the functional determinant $ \det ( \mathcal{L}_{0,ij} ( \Phi ))^{-1}$, where $ \mathcal{L}_{0} = \frac{1}{2}  ( \partial_{\mu} \phi )^{2} - g \phi^{3} /3!$. The combinatorial factors associated with the diagrams of Fig.~\ref{fig:23scalar} account for doubling of the usual one-loop results that follows from $ \mathcal{L}_{0} $ alone.


In Refs. \cite{Brandt:2018lbe, Brandt:2019ymg, McKeon:1992rq,Brandt:2021qgh, Brandt:2020gms, Brandt:2021nev} the approach we have just illustrated with a simple scalar model is applied to Yang-Mills and gravitational theory. There it is shown that just as in this simple scalar model, the effect of the Lagrange multiplier field is to reproduce the usual one-loop effects and to eliminate all effects beyond one-loop order. Since unitarity and gauge invariance are retained, only the usual transverse degrees of freedom propagate in these gauge theories when supplemented by a Lagrange multiplier field in the way we have described. Renormalization of these models is also discussed in Refs. \cite{Brandt:2018lbe, Brandt:2019ymg, McKeon:1992rq,Brandt:2021qgh, Brandt:2020gms, Brandt:2021nev}.
The additional ghost fields introduced in Section II serve to ensure that the path integral is invariant under a change of variables and that loop calculations coincide exactly with just the usual one loop results that follow from the classical Lagrangian alone (for more details, see Ref. \cite{Brandt:2022kjo}).

\subsection{Yang-Mills}\label{section:YMpELM}

\begin{figure}[ht]
    \includegraphics[scale=0.8]{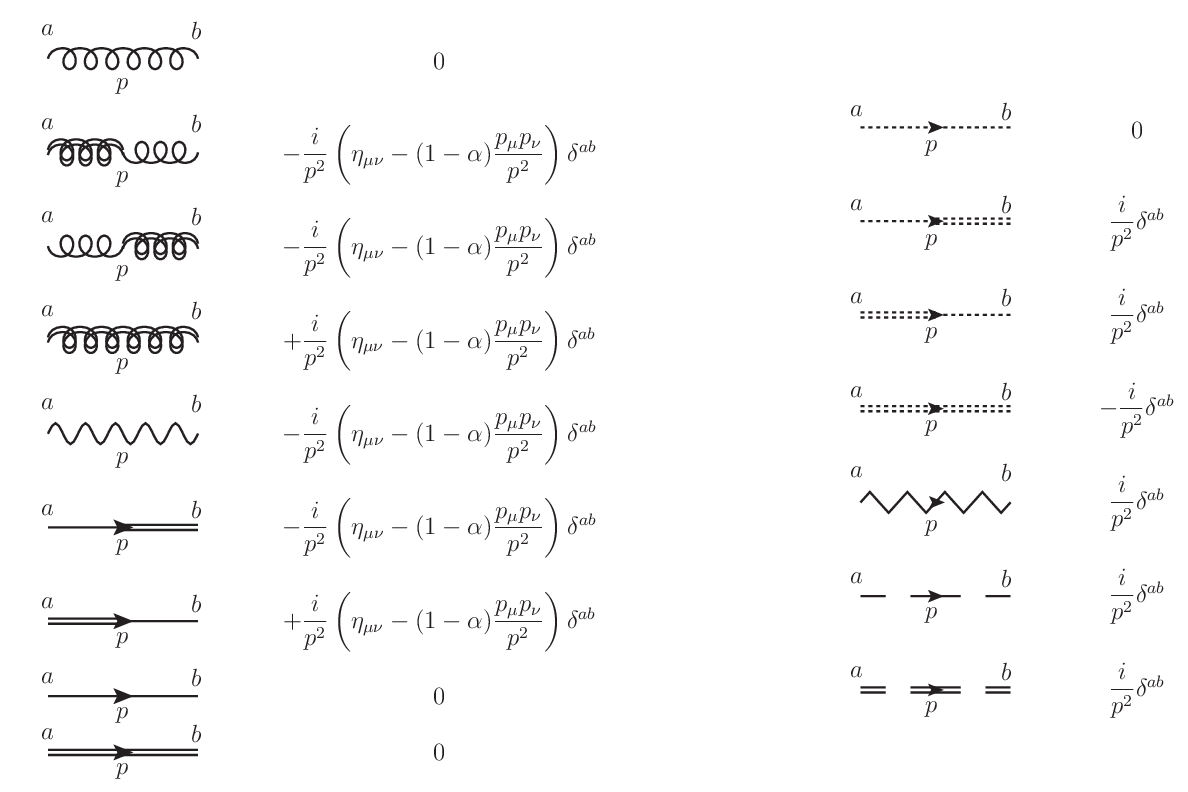}
    \caption{Propagators of the Yang-Mills theory in the extended Lagrange multiplier formalism. The gauge fields $ A_{\mu}^{a} $, $ \lambda_{\mu}^{a} $, $ \psi_{\mu}^{a} $, $ \theta_{\mu}^{a} $ and $ \chi_{\mu}^{a} $ are respectively represented by spring, double spring, solid, double solid, and wavy lines. The associated Faddeev-Popov ghost fields $c^{a} $ ($ \bar{c}^{a}$), $ d^{a} $ ($ \bar{d}^{a}$), $ e^{a} $ ($ \bar{e}^{a}$), $ \gamma^{a} $ ($ \tilde{\gamma}^{a}$), and $ \epsilon^{a} $ ($ \tilde{\epsilon}^{a}$) by dotted, double dotted, zigzag, dashed, and double dashed lines. }\label{fig:propYM}
\end{figure}

\begin{figure}[ht]
    \includegraphics[width=1\textwidth]{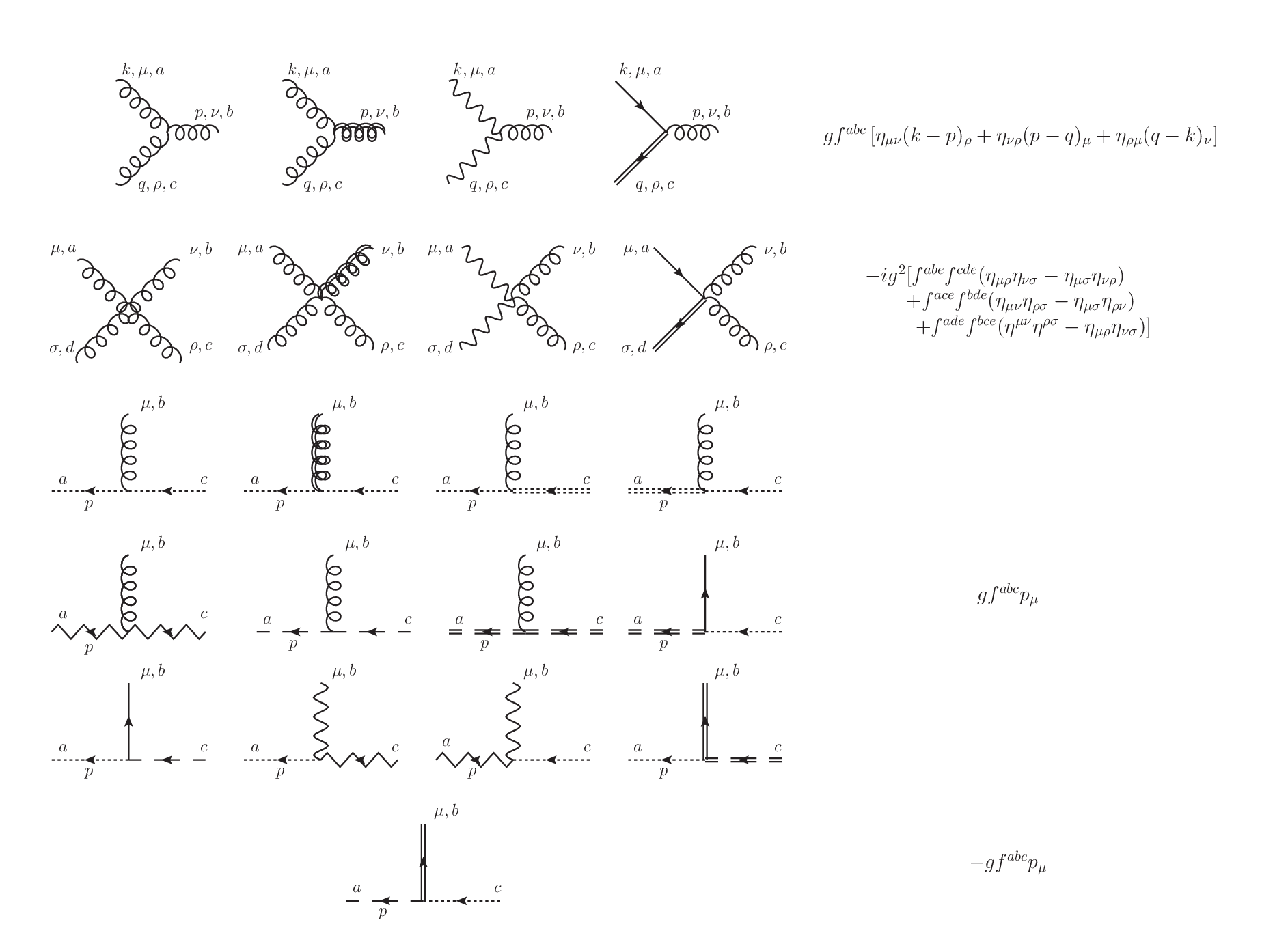}
    \caption{Vertices of the extended Yang-Mills theory obtained from the action in Eq.~\eqref{eq:611}.}\label{fig:vertYM}
\end{figure}

Here, we will derive the propagators of the Yang-Mills theory in the framework of the extended Lagrange multiplier theory studied in Section V. Let us organize the bilinear terms in the fields in Eq.~\eqref{eq:611}  as 
\begin{equation}\label{eq:appB10}
    \begin{split}
        \mathfrak{L}^{(2)} ={}& \frac{1}{2} \begin{pmatrix}
        A_{\mu}^{a} &  \lambda_{\mu}^{a} 
    \end{pmatrix}
    \begin{pmatrix}
        a_{\mu \nu}^{ab} & a_{\mu \nu}^{ab} \\ 
        a_{\mu \nu}^{ab} & 0
    \end{pmatrix}
    \begin{pmatrix}
        A_{\nu}^{b} \\ \lambda_{\nu}^{b}  
    \end{pmatrix}
    +  \frac{1}{2} \begin{pmatrix}
        \chi_{\mu}^{a} & \psi_{\mu}^{a} & \theta_{\mu}^{a}
    \end{pmatrix}
    \begin{pmatrix}
        a_{\mu \nu}^{ba} & 0 & 0\\
        0 & 0 & a_{\mu \nu}^{ab} \\
        0 & -a_{\mu \nu}^{ab} & 0 \\
    \end{pmatrix}
    \begin{pmatrix}
        \chi_{\nu}^{b} \\
        \psi_{\nu}^{b} \\
        \theta_{\nu}^{b}
    \end{pmatrix}
    \\
                              & + \begin{pmatrix}
                                  \bar{c}^{a} & \bar{d}^{a} & \bar{e}^{a} & \tilde{\gamma}^{a} & \tilde{\epsilon}^{a} 
                              \end{pmatrix}
                              \begin{pmatrix}
                                  a^{ab} & a^{ab} & 0 & 0 & 0 \\
                                  a^{ab} & 0 & 0 & 0 & 0 \\
                                  0 & 0 & a^{ab} & 0 & 0 \\
                                  0 & 0 & 0 & a^{ab} & 0 \\
                                  0 & 0 & 0 & 0 & a^{ab}  
                              \end{pmatrix}
\begin{pmatrix}
                                  c^{b} \\d^{b} \\ e^{b} \\ \gamma^{b} \\ \epsilon^{b} 
                              \end{pmatrix},
        \end{split}
\end{equation}
where 
\begin{equation}\label{eq:appB11}
    a_{\mu \nu}^{ab} = \left [ \eta_{\mu \nu} \partial^{2} - \left ( 1 - \frac{1}{\alpha}\right ) \partial_{\mu} \partial_{\nu} \right ] \delta^{ab} \quad \text{and} \quad a^{ab} = - \partial^{2} \delta^{ab} . 
\end{equation}
To obtain the propagators, we use that 
\begin{equation}\label{eq:appB14}
\begin{pmatrix}
    a & a \\
    a & 0
\end{pmatrix}^{-1}
= \begin{pmatrix}
    0 & a^{-1}\\
    a^{-1} & - a^{-1}
\end{pmatrix}
 \quad \text{and} \quad \begin{pmatrix}
     0 & a \\
     -a & 0
 \end{pmatrix}^{-1} = \begin{pmatrix}
     0 & -a^{-1} \\
     a^{-1} & 0
 \end{pmatrix}.
\end{equation}
Note that the inverse of $ -i a_{\mu \nu}^{ab} $ yields the usual propagator of the gauge field in Yang-Mills theory
\begin{equation}\label{eq:appB12}
    - \frac{i}{p^{2}} \left ( \eta_{\mu \nu} - (1 - \alpha ) \frac{p_{\mu} p_{\nu}}{p^{2}}\right ) \delta^{ab},
\end{equation}
and the inverse of $ -i a^{ab} $ yields the associated ghost field propagator (in the Lorenz gauge):
\begin{equation}\label{eq:appB13}
    \frac{i}{p^{2}} \delta^{ab} .
\end{equation}
We present the Feynman rules for the propagators and the vertices in Figs.~\ref{fig:propYM} and~\ref{fig:vertYM}.

From Eqs.~\eqref{eq:appB10} and \eqref{eq:appB14}, we see that there is no propagator $ \langle 0|T A_{\mu}^{a} A_{\nu}^{b}| 0 \rangle $ for the gauge field $ A_{\mu}^{a} $. However, there are now mixed propagators $ \langle 0|T \lambda_{\mu}^{a} A_{\nu}^{b}| 0 \rangle = \langle 0|T A_{\mu}^{a} \lambda_{\nu}^{b}| 0 \rangle$ and a propagator $ \langle 0|T \lambda_{\mu}^{a} \lambda_{\nu}^{b}| 0 \rangle $ for the Lagrange multiplier field $ \lambda_{\mu}^{a} $. 
Since there is no propagator for $A_{\mu}^{a} $, and $ \lambda_{\mu}^{a}$ only enter linearly in any vertex (see Fig.~\ref{fig:vertYM}), it is impossible to draw a one-particle irreducible diagram with two or more external fields $ \lambda_{\mu}^{a} $. 

Moreover, since unitarity and gauge invariance are retained, only the usual transverse degrees of freedom propagate in the Yang-Mills theory when supplemented by a Lagrange multiplier field as described in Section V. The number of degrees of freedom of the Yang-Mills theory in the framework of the extended Lagrange multiplier formalism may be inferred from Eq.~\eqref{eq:611} as 
\begin{equation}\label{eq:appB15}
    N_{\text{eYM}} = N_{A} + N_{ \lambda} + N_{\chi} + N_{\gamma} + N_{\epsilon} - N_{\text{gh}}, 
\end{equation}
where the fields $c$ ($ \bar{c}$), $ d$ ( $ \bar{d} $), $ e $ ($ \bar{e} $), $ \psi $, $ \theta$ serves as negative degrees of freedom\footnote{These fields $c$ ($ \bar{c}$), $ d$ ( $ \bar{d} $), $ e $ ($ \bar{e} $), $ \psi $, $ \theta$ would violate the spin-statistics theorem.}: $ N_{ \text{gh} } = N_{c} + N_{d} + N_{e} + N_{\psi} + N_{\theta} $. Since  $ N_{A} = N_{\lambda} = N_{\chi} = N_{\psi} = N_{\theta}   $, and $  N_{c} = N_{d} = N_{e} =N_{\gamma} = N_{\epsilon} $, we have that 
\begin{equation}\label{eq:appB16}
    N_{ \text{eYM} } = N_{A} - N_{c} = 2 \mathop{\rm Dim} G
\end{equation}
which is the number of degrees of freedom of the standard Yang-Mills theory with gauge group $G$.

\section{Gauge algebra of the extended Lagrange multiplier theory}\label{ssection:BRST}

Consider that the gauge transformations \eqref{eq:9} form a closed algebra. Thus, the (super)commutator of two gauge transformations of the form of Eq.~\eqref{eq:9} must itself be a gauge transformation:
\begin{equation}\label{eq:GF:supercommutator}
    \left(\delta_{1} \delta_{2} - \delta_{2} \delta_{1} \right)_{\xi}  \phi_{i} = \left ( H_{im,l} H_{ln} - H_{in,l}H_{lm} \right ) \xi_{2m} \xi_{1n} \equiv \tensor{f}{_{mn|p}}  H_{ip} \xi_{2m} \xi_{1n} = \delta_{\xi_{3}} \phi_{i},
\end{equation}
where $ \xi_{3 \, p} = \tensor{f}{_{mn}_{|p}} \xi_{2 \, m} \xi_{1 \, n} $. 
We also assume that the conditions \eqref{eq:34} hold. 
The structure constants satisfy 
 \begin{equation}\label{eq:GF:StructureConstantsJacobi}
     \tensor{f}{_{qa}_{|p}} 
     \tensor{f}{_{bc}_{|q}} + \tensor{f}{_{qc}_{|p}} \tensor{f}{_{ab}_{|q}} + \tensor{f}{_{qb}_{|p}} \tensor{f}{_{ca}_{|q}} =0
 \end{equation}
 which leads to the Jacobi identity
\begin{equation}\label{eq:GF:JacobiId}
     [[\delta_{1} , \delta_{2} ] , \delta_{3} ]+ 
     [[\delta_{2} , \delta_{3} ] , \delta_{1} ]+ 
     [[\delta_{3} , \delta_{1} ] , \delta_{2} ]=0.
 \end{equation}

 We note that Eq.~\eqref{eq:GF:supercommutator} holds for any field. For example, the differentiation of Eq.~\eqref{eq:33}  with respect to $ \phi_{m} $ implies that 
\begin{equation}\label{eq:BB1}
    \left ( H_{ij , k} H_{kl,m} - H_{il , k} H_{kj,m}\right )  = f_{jl|k} H_{ik,m} ,
\end{equation}
then we find that 
\begin{equation}\label{eq:GF:supercommutator1lambda}
    [ \delta_{1} , \delta_{2} ]_{\xi} \lambda_{i} = \tensor{f}{_{mn}_{|p}} H_{ip, k} \lambda_{k} \xi_{2 \, m} \xi_{1 \, n},
\end{equation}
which can be rewritten as
\begin{equation}\label{eq:GF:supercommutator1lambdab}
    [ \delta_{1} , \delta_{2} ]_{\xi} \lambda_{i} = H_{ip, k} \lambda_{k} \xi_{3 \, p} = \delta_{\xi_{3}} \lambda_{i}.
\end{equation}
Similar relations can be derived for the ghost fields $ \chi_{i} $, $ \psi_{i} $ and $ \theta_{i} $.

This shows that the extended Lagrange multiplier formalism is consistent with the gauge algebra of the theory without Lagrange multiplier fields: 
\begin{equation}\label{eq:GF:mLagrangemultiplier:commutatorgt0}
    [ \delta_{\xi} , \delta_{\xi} ] = \delta_{\xi}.
\end{equation}
Then, by using Eq.~\eqref{eq:33} and Eq.~\eqref{eq:BB1}, one finds that 
 \begin{equation}\label{eq:algebraextended}
     \begin{split}
         & [ \delta_{1}, \delta_{2} ]_{\zeta} = \delta_{\zeta_{3}} , \quad  [ \delta_{\zeta} , \, \delta_{\zeta} ] = 0,  \quad [ \delta_{1} , \, \delta_{2} ]_{\sigma} = \delta_{\zeta_{31}} ,  \quad \{ \delta_{\pi} , \, \delta_{\pi} \} = 0,   \quad \{ \delta_{\tau} , \, \delta_{\tau} \} = 0,  \\ 
 & [ \delta_{\xi} , \, \delta_{\zeta} ] = \delta_{\zeta_{32} },   \quad [ \delta_{\xi} , \, \delta_{\sigma} ] = \delta_{\sigma_{3} },    \quad [ \delta_{\xi} , \, \delta_{\pi} ] = \delta_{\pi_{3} },   \quad [ \delta_{\xi } , \, \delta_{\tau} ] = \delta_{\tau_{3} },  \\ 
 & [ \delta_{\zeta} , \, \delta_{\sigma} ] = 0,   \quad [ \delta_{\zeta} , \, \delta_{\pi} ] = 0,  \quad [ \delta_{\zeta} , \, \delta_{\tau} ] = 0, \\ 
 & [ \delta_{\sigma} , \, \delta_{\pi} ] = 0,   \quad [ \delta_{\sigma} , \, \delta_{\tau} ] = 0,  \\ 
 & \{\delta_{\pi} , \, \delta_{\tau} \} = \delta_{\zeta_{33} };
     \end{split}
 \end{equation}
 where $ \zeta_{3 \, k} = f_{ij|k} \xi_{1 \, i} \xi_{2 \, j} $, $ \zeta_{31 \, k} = f_{ij|k} \sigma_{1 \, i} \sigma_{2 \, j} $, $ \zeta_{32 \, k} = f_{ij|k} \zeta_{i} \xi_{j} $, $ \sigma_{3 \, k} = f_{ij|k} \sigma_{i} \xi_{j} $, $ \pi_{3 \, k} = f_{ij|k} \pi_{i} \xi_{j} $, $ \tau_{3 \, k} = f_{ij|k} \tau_{i} \xi_{j} $  and $ \zeta_{33 \, k} = - f_{ij | k}  \tau_{i} \pi_{j} $.

By direct calculations, one also can show that all the generators of the extended gauge algebra satisfy the Jacobi identity in Eq.~\eqref{eq:GF:JacobiId}. 
We have, for example, that 
\begin{equation}\label{eq:BB8}
    [[\delta_{\xi_{1} } , \delta_{\xi_{2} } ] , \delta_{\sigma} ] 
+ 
[[\delta_{\xi_{2} } , \delta_{\sigma} ] , \delta_{\xi_{1} } ] 
+
[[\delta_{\sigma} , \delta_{\xi_{1} } ] , \delta_{\xi_{2} } ] 
\sim f_{ij|k} f_{kn|l} + f_{jn|k} f_{ki|l} - f_{in|k} f_{kj|l}
\end{equation}
which by Eq.~\eqref{eq:GF:StructureConstantsJacobi} must vanish.
Thus, the gauge algebra of the extended Lagrange multiplier theory in Eq.~\eqref{eq:algebraextended} is closed. We find that the extended Lagrange multiplier theory is also a YM type theory. Moreover, Eq.~\eqref{eq:algebraextended} is consistent with the results in Eqs.~\eqref{eq:36} and \eqref{eq:39}.

\section{A superdeterminant identity}\label{section:YMtLagrangemultiplier}

Defining the block matrix
\begin{equation}\label{B1}
    \begin{pmatrix}
        \bm{P} & \bm{Q} \\
        \bm{R} & \bm{S}
    \end{pmatrix}
    =
\begin{pmatrix}
    0 & A & 0 & 0 & 0 \\ 
    A & B & C  & -D & E  \\
    0 & C & A & 0 & 0 \\
    0 & E  & 0 & A & 0 \\
    0 & D & 0 & 0 & A \\ 
    \end{pmatrix}, 
\end{equation}
in which we identify 
\begin{equation}\label{eq:B2}
    \bm{P} = \begin{pmatrix}
        0 & A \\
        A & B
    \end{pmatrix}, \quad 
    \bm{Q} = \begin{pmatrix}
        0 & 0 & 0 \\
        C & -D & E 
    \end{pmatrix}, \quad 
   \bm{R} = \begin{pmatrix}
       0 & C \\
       0 & E \\
       0 & D 
    \end{pmatrix}, \quad 
    \bm{S} = \begin{pmatrix}
        A & 0 & 0 \\
        0 & A & 0 \\
        0 & 0 & A \\
    \end{pmatrix}.
\end{equation}
Using the identity 
\begin{equation}\label{eq:B3}
 \begin{pmatrix}
        \bm{P} & \bm{Q} \\
        \bm{R} & \bm{S}
        \end{pmatrix} =  \begin{pmatrix}
        \bm{1} & \bm{Q} \\
        0 & \bm{S}
    \end{pmatrix}
    \begin{pmatrix}
        \bm{P} - \bm{Q} \bm{S}^{-1} \bm{R} & 0 \\
        \bm{S}^{-1} \bm{R} & \bm{1}
    \end{pmatrix}
\end{equation}
and that 
\begin{equation}\label{eq:B3a}
    \mathop{\rm Sdet}(\bm{P} - \bm{Q} \bm{S}^{-1} \bm{R}) = \mathop{\rm Det} \begin{pmatrix}
        0 & A \\
        A & B - F
    \end{pmatrix}
    = \mathop{\rm Det} \bm{P},
\end{equation}
where $ F = C A^{-1} C - D A^{-1} E + E A^{-1} D $,
implies that
\begin{equation}\label{eq:B5}
    \mathop{\rm Sdet} \begin{pmatrix}
        \bm{P} & \bm{Q} \\
        \bm{R} & \bm{S}
        \end{pmatrix} = \mathop{\rm Sdet} \bm{S}  \mathop{\rm Det} \bm{P} = \det A,
\end{equation}
which demonstrates the result in Eq.~\eqref{eq:ghostdettotal}.

\section{Zinn-Justin master equation}\label{section:BRSTZJ}

Introducing sources for the fields in Eq.~\eqref{eq:25} and Eq.~\eqref{eq:26} leads to the following generating functional for the extended Lagrange multiplier theory  
\begin{equation}\label{eq:616D}
    Z [ \bm{J} ,\bm{\eta} , \bar{\bm{\eta}}]   = 
    \int \mathop{\mathcal{D} \bm{\phi}_{i}} \mathop{\mathcal{D} \bm{B}_{i}} \mathop{\mathcal{D} \bm{c}_{i} } \mathop{\mathcal{D} \bar{\bm{c}}_{i} } 
    \exp i \int \mathop{d x} \left ( 
        \mathcal{L}_{ \text{T} } 
        + \bar{\bm{J}}_{i} \bm{\phi}_{i} + \bar{\bm{\eta}}_{i} \bm{c}_{i} +  \bar{\bm{c}}_{i} \bm{\eta}_{i}  
\right ), 
\end{equation}
where $ \bm{\phi}_{i} = ( \phi_{i} , \lambda_{i} , \chi_{i} , \psi_{i} , \theta_{i} )$, $ \bm{B}_{i} = (B_{i} , E_{i} , G_{i} , \Xi_{i} , \Omega_{i} )$, $ \bm{c}_{i} = ( c_{i} , d_{i} , e_{i} , \gamma_{i} , \epsilon_{i} )$ and the Lagrangian $ \mathcal{L}_{\text{T}} $ is defined in Eq.~\eqref{eq:27}. 

Following Refs. \cite{Taylor:1971ff, Slavnov:1972fg, Kluberg-Stern:1974iel}, we also introduce in Eq.~\eqref{eq:616D} sources for the non-linear BRST variations (the BRST variation $s A$ is defined by  $ \delta A = (s A) \eta $): the source $ \bm{U}_{i} $ to the variation of the field $ s\bm{\phi}_{i}  $ and $ \bar{\bm{V}}_{i} $ to the variation of the ghost field $ s\bm{c}_{i} $.  We obtain  
\begin{equation}\label{eq:616}
    Z [ \bm{J}  ,\bm{\eta} ,  \bar{\bm{\eta}}; \bm{U}, \bm{V} ]   = 
    \int \mathop{\mathcal{D} \bm{\phi}_{i}} \mathop{\mathcal{D} \bm{c}_{i} } \mathop{\mathcal{D} \bar{\bm{c}}_{i} } 
    \exp i \int \mathop{d x} \left ( 
        \mathcal{L}_{ \text{T} } 
        + \bar{\bm{J}}_{i} \bm{\phi}_{i}  + \bar{\bm{\eta}}_{i} \bm{c}_{i} +  \bar{\bm{c}}_{i} \bm{\eta}_{i}  
    + \bm{U}_{i} s \bm{\phi}_{i} + \bar{\bm{V}}_{i} s \bm{c}_{i} 
\right ).
\end{equation}

The invariance of Eq.~\eqref{eq:616} under the BRST transformations implies that 
\begin{equation}\label{eq:618}
    \begin{split} 
        \delta Z [ \bm{J}  , \bm{\eta} , \bar{\bm{\eta}}; \bm{U}, \bm{V} ]   
        ={}& 
    \int \mathop{\mathcal{D} \bm{\phi}_{i}} \mathop{\mathcal{D} \bm{c}_{i} } \mathop{\mathcal{D} \bar{\bm{c}}_{i} } \left [ i \int \mathop{d x} \left(
     \bar{\bm{J}}_{i} \delta \bm{\phi}_{i} + \bar{\bm{\eta}}_{i} \delta \bm{c}_{i} +  \delta \bar{\bm{c}}_{i}  \bm{\eta}_{i}  
\right)\right ] \\ & \times  
\exp i \int \mathop{d x} \left ( 
        \mathcal{L}_{ \text{T} } 
    + \bar{\bm{J}}_{i} \bm{\phi}_{i} + \bar{\bm{\eta}}_{i} \bm{c}_{i} +  \bar{\bm{c}}_{i} \bm{\eta}_{i}  
    + \bm{U}_{i} \delta \bm{\phi}_{i} + \bar{\bm{V}}_{i} \delta \bm{c}_{i} 
\right )=0, 
    \end{split}
\end{equation}
where we used that $ \mathcal{L}_{\text{T}} +  \bm{U}_{i} \delta \bm{\phi}_{i} + \bar{\bm{V}}_{i} \delta \bm{c}_{i} $ is invariant under BRST transformations, since $ \delta s \bm{U}_{i} = \delta s \bar{\bm{V} }_{i} = 0$ which follows from the nilpotency of the BRST transformations ($ s^{2} =0$).

Using Eq.~\eqref{eq:30} and 
\begin{subequations}\label{eq:619}
    \begin{align} 
    \frac{\delta_{L}  
    Z [ \bm{J} ,\bm{\eta} , \bar{\bm{\eta}}; \bm{U}, \bm{V} ]   }{\delta \bm{U}_{i}} 
    ={}& s\bm{\phi}_{i}, \\
    \frac{\delta_{L}           Z [ \bm{J}  , \bm{\eta} , \bar{\bm{\eta}}; \bm{U}, \bm{V} ]   }{\delta \bm{V}^{a }} 
    ={}& s\bm{c}_{i}, \\
\end{align}  
\end{subequations}
we find that
\begin{equation}\label{eq:620}
    \int \mathop{d x}  
    \left(\bar{\bm{J}}_{i}  
        \frac{ \delta_{L} Z [ \bm{J}  , \bm{\eta} , \bar{\bm{\eta}}; \bm{U}, \bm{V} ]}{\delta \bm{U}_{i}} 
    + \bar{\bm{\eta}}_{i} 
    \frac{\delta_{L}  Z [ \bm{J}  , \bm{\eta} , \bar{\bm{\eta}}; \bm{U}, \bm{V} ]   }{\delta \bm{V}_{i}} 
    + s \bar{\bm{c}}_{i}     
    \bm{\eta}_{i} 
\right) \eta 
    =0,
\end{equation}
where the subscript $L$ ($R$) denotes left (right) differentiation. 

We can write Eq.~\eqref{eq:620} into a relation in terms of the generating functional $ \Gamma $ of one-particle irreducible Green's functions. This can be done by using the Legendre transform of the generating functional of connected Green's function $ W [ \bm{J} ,  \bm{\eta} , \bar{\bm{\eta}}; \bm{U}, \bm{V} ] \equiv -i \ln  Z [ \bm{J} ,   \bm{\eta} , \bar{\bm{\eta}}; \bm{U}, \bm{V} ]$:
\begin{equation}\label{eq:621}
    W [  \bm{J} , \bm{\eta} , \bar{\bm{\eta}}; \bm{U}, \bm{V} ] = 
    \Gamma [ \bm{\phi} , \bm{B} ,\bm{c} , \bm{U} , \bm{V} ] 
    + \int \mathop{d^{4} x} \left (
        \bar{\bm{J}}_{i} \bm{\phi}_{i}  + \bar{\bm{\eta}}_{i} \bm{c}_{i} +  \bar{\bm{c}}_{i} \bm{\eta}_{i}  
    \right ) 
\end{equation}
Thus, we obtain the Zinn-Justin master equation \cite{Zinn-Justin:2011bia} 
\begin{equation}\label{eq:622}
    \int \mathop{d^{4}x} \left(\frac{\delta_{R} \Gamma }{\delta \bm{\phi}_{i} } \frac{\delta_{L} \Gamma }{\delta \bm{U}_{i}  }+ \frac{\delta_{R} \Gamma }{\delta \bm{c}_{i}} \frac{\delta_{L} \Gamma}{\delta \bm{V}_{i}} -  s\bar{\bm{c}}_{i}   \frac{\delta_{R} \Gamma }{\delta \bar{ \bm{c} }_{i}}\right) 
    =0. 
\end{equation}

The so-called Slavnov-Taylor identities \cite{Taylor:1971ff, Slavnov:1972fg} can be derived by taking functional derivatives of Eq.~\eqref{eq:622} with respect to the fields.  These identities are mainly used in the study of the renormalizability of gauge theories. However, they may also be used to derive relations between proper Green's functions which may be relevant. 

Introducing a source $ \bm{N}_{i} $ to $ s\bar{\bm{c}}_{i} $ in Eq.~\eqref{eq:616D} allows us to rewrite the master equation \eqref{eq:622} as 
\begin{equation}\label{eq:624}
    ( \Gamma , \Gamma ) = 0,
\end{equation}
where $( \cdot , \cdot )$ is the Batalin-Vilkovisky anti-bracket \cite{Batalin:1981jr}
\begin{equation}\label{eq:623}
    ( A , B ) \equiv 
    \int \mathop{d^{4}x} \left(
        \frac{\delta_{R} A }{\delta \bm{\Phi}_{i} } \frac{\delta_{L} B }{\delta \bm{\Phi}_{i}^{*}   }- 
        \frac{\delta_{R} A }{\delta \bm{\Phi}_{i}^{*}  } \frac{\delta_{L} B }{\delta \bm{\Phi}_{i}  }
    \right), 
\end{equation}
where $ \bm{\Phi}_{i} = ( \bm{\phi}_{i} , \bm{B}_{i}, \bm{c}_{i} , \bar{\bm{c}}_{i} )$ are the fields and $ \bm{\Phi}_{i}^{*} =( \bm{U}_{i} , \bm{N}_{i} , \bm{V}_{i} )$ are known as the anti-fields (of opposite statistics to the $ \bm{\Phi}_{i} $). The Batalin-Vilkovisky formalism \cite{Batalin:1981jr, Batalin:1983ggl} is widely used to prove the gauge invariant renormalizability of gauge theories \cite{Zinn-Justin:1974ggz, Lavrov:2012xm, Lavrov:2022ceg}.

\bibliography{all.bib}

\end{document}